\documentclass[
reprint,
amsmath,
amssymb,
aps,
prx,
superscriptaddress,
longbibliography
]{revtex4-2}

\usepackage{graphicx}
\usepackage{bm}
\usepackage{booktabs}
\usepackage[colorlinks=false, pdfborder={0 0 0}]{hyperref}
\usepackage{orcidlink}
\begin{document}

\title{Bayesian Variational Method for Precision Few-Body Calculations}

\author{Shigeyoshi Aoyama\,\orcidlink{0000-0002-3784-4544}}
\email{shigeyoshi.aoyama@kek.jp}
\affiliation{Computing Research Center, High Energy Accelerator Research
Organization (KEK), \\ Tsukuba, Ibaraki 305-0801, Japan}

\begin{abstract}
Many variational descriptions of quantum many-body systems rest on an
expansion over basis functions, and their practical limit is often set
by the number of basis functions required. We propose the Bayesian
variational method (BVM), in which the basis functions are selected by
Bayesian optimization: a Gaussian-process surrogate model, conditioned on
the candidates evaluated so far, predicts which candidates are most likely
to lower the energy, and the candidate evaluations, being mutually
independent, are distributed over many nodes. Two further ingredients make the method practical. An
incremental diagonalization, adapted here to the present setting,
evaluates each candidate by reusing the previous diagonalization of the
accepted basis instead of solving the full eigenvalue problem anew. A trimming procedure, employed in this paper, continually removes
basis functions that have become nearly linearly dependent, keeping the
number of accepted basis functions small while guiding the basis toward the optimal solution. The
BVM is broadly applicable to energy variational problems based on
basis-function expansions in quantum mechanics; in this work we apply it
to the Gaussian expansion method (GEM), one of the standard approaches
in few-body physics. Because the GEM uses a nonorthogonal basis set, its
linear dependence is strong, so that the basis reduction achieved by the
BVM is large. In this application, the reduction of the basis both
accelerates the computation and, more importantly, greatly reduces the
memory requirement---one of the central bottlenecks of the variational
method: the reference energy of the full $32{,}000$-dimensional GEM
diagonalization is reproduced to within $0.01$\,K with only $705$ basis
functions and to within $0.001$\,K with $2{,}127$, corresponding to
memory reductions of $99.95\%$ and $99.56\%$, since the matrix storage
grows as the square of the basis dimension. Within the GEM, this opens
a path to the precision study of six- and seven-body systems, and
beyond, that has so far been difficult to reach.
\end{abstract}

\maketitle

\section{Introduction}
\label{sec:intro}

The variational method, founded on the Ritz principle~\cite{Ritz1909}, is one of the principal approaches to the quantum-mechanical many-body problem. It was introduced in the early quantum-mechanical treatments of the helium atom~\cite{Kellner1927,Hylleraas1928} as an efficient means of determining the wave function, and it remains a representative method to this day. In this method one assumes a trial wave function with adjustable parameters and searches for the parameters that lower the energy. Energy variation is, in essence, an optimization problem: the search for the parameters that realize the state of lowest energy.

In recent years, machine learning and artificial intelligence have come into broad use throughout society. Among the methods founded on Bayes' theorem~\cite{Bayes1763}, Bayesian inference~\cite{MacKay1992} and Bayesian machine learning~\cite{Gelman2013,Murphy2012} are very widely used. In the natural sciences, for example, Gaussian-process regression (GPR)~\cite{WilliamsRasmussen1996} serves to analyze observed experimental data and to predict unobserved data. Here we bring these ideas to bear on the quantum-mechanical variational method.

In the variational method a trial wave function $\psi(\bm{\theta})$ is specified by parameters $\bm{\theta}$, and its energy is the Rayleigh quotient
\begin{equation}
E(\bm{\theta})=\frac{\langle\psi(\bm{\theta})|H|\psi(\bm{\theta})\rangle}
{\langle\psi(\bm{\theta})|\psi(\bm{\theta})\rangle},
\end{equation}
an upper bound to the exact energy. The task is the minimization
\begin{equation}
\bm{\theta}_{\rm opt}=\operatorname*{arg\,min}_{\bm{\theta}} E(\bm{\theta}).
\end{equation}
We cast this search as Bayesian estimation. Writing $\mathcal{E}=\{E_1,E_2,\ldots,E_k\}$ for the $k$ lowest eigenenergies obtained by diagonalization, the parameters are estimated through Bayes' theorem,
\begin{equation}
P(\bm{\theta}\mid \mathcal{E})=\frac{P(\mathcal{E}\mid\bm{\theta})\,P(\bm{\theta})}{P(\mathcal{E})},
\end{equation}
the optimum being the $\bm{\theta}$ favored by the posterior $P(\bm{\theta}\mid\mathcal{E})$. The variational search for the energy minimum is thereby recast as Bayesian estimation of the parameters from the evaluated energies.

The parameters $\bm{\theta}$ are of two kinds. In many cases the trial wave function is taken as a superposition of basis functions, and for a fixed set of basis functions the superposition coefficients follow from a linear generalized eigenvalue problem: the Hamiltonian and overlap matrices are constructed in the basis and diagonalized, yielding the energy and the wave function. These coefficients therefore call for no separate search. The remaining parameters---those that specify the basis functions themselves---are far more demanding to determine. When the basis functions are drawn from a discretized set of candidates, their selection is a combinatorial optimization problem, and it is to this selection that the Bayesian estimation is directed: the aim is to represent the wave function to a prescribed accuracy with as few basis functions as possible.

To appreciate the scale of this combinatorial problem, consider the ground state alone. The number of basis functions in the superposition is in principle unbounded, but the energy converges once their number reaches some $N_{\rm conv}$. If each continuous parameter of a basis function is restricted to a finite grid and the functional form is fixed, the product of the grid sizes gives the total number of candidate functions, $N_{\rm total}$. Choosing the $N_{\rm conv}$ that lower the energy the most is then a combinatorial optimization over ${}_{N_{\rm total}}C_{N_{\rm conv}}$ possibilities---a number so large that the problem lies beyond the reach of even the most advanced supercomputers.

Since this combinatorial optimization cannot be carried out exhaustively, the basis has to be assembled by some tractable prescription. The most immediate one is to grow it incrementally, examining the candidate functions one at a time and retaining each only when it lowers the energy. This is the guiding idea of the stochastic variational method (SVM)~\cite{Varga95,textbook}, a method aimed at reaching the lowest-energy solution with as few basis functions as possible, in which the candidates are drawn at random and admitted whenever they improve the ground-state energy. The SVM has served as a very powerful tool in few-body physics and has been employed by many researchers. Its acceptance is decided one candidate at a time, on the immediate lowering of the energy alone, and thus does not perform the combinatorial optimization posed above: each function is weighed only against the basis present at that moment, so that a function admitted early may be rendered redundant by those accepted after it, and the resulting basis carries considerable redundancy, exceeding the $N_{\rm conv}$ of an optimal selection. Developments along this line, though likewise not founded on Bayesian optimization, include SVM+~\cite{Aoyama2025}, a parallel SVM~\cite{base}, and gradient variational methods (GVM)~\cite{Recchia2024}.

In this paper we propose the Bayesian variational method (BVM), in which the selection of the basis functions is carried out by Bayesian optimization. Selecting the necessary functions from the full candidate set of size $N_{\rm total}$ at once is out of reach; instead, at each step a pool of candidates is drawn at random from $N_{\rm total}$, and the basis is enlarged by a group of $K$ functions---rather than a single one---obtained by optimizing the selection within this pool. Enlarging the basis in groups is also expected to be more efficient whenever the correlations among basis functions are strong. A typical example is the tensor correlation in nuclei: in the deuteron ground state, the $L=0$ and $L=2$ components couple strongly through the off-diagonal matrix elements and thereby gain a large binding energy. In such cases the simultaneous adoption of several functions, as in the present BVM, can treat these off-diagonal components directly, in contrast to the one-by-one adoption of the conventional SVM. The basis so built is not the exact optimum over $N_{\rm conv}$ functions; instead, as a result of the optimization at each step, a small basis close to $N_{\rm conv}$ is obtained. A surrogate model---here GPR, one of the standard choices---is conditioned on the energies evaluated so far and returns, for every candidate, a predicted energy together with an uncertainty. An acquisition function combines the two into a single criterion, favoring candidates whose predicted energy is low while still probing those whose energy remains uncertain, so that a promising candidate hidden behind a large uncertainty is not overlooked; the concrete forms of the surrogate model and of the acquisition function are given in Sec.~\ref{sec:method}.

Each explicit evaluation would ordinarily require a diagonalization; to avoid this, we develop an incremental diagonalization that reuses the diagonalization already performed for the accepted basis---its stored matrix factorization and ground-state vector---and reduces the cost of each candidate to $O(N^2)$. Since the candidate evaluations are mutually independent, they are distributed over many nodes without communication. A trimming procedure, applied as the basis grows, removes functions that have become nearly linearly dependent and keeps the number of accepted basis functions small.

To validate the proposed method, we adopt as a benchmark a typical few-body system, the $^4$He pentamer, described by the Gaussian expansion method (GEM)~\cite{Hiya2003}, a standard method for few-body systems. We use the corresponding Hamiltonian and overlap matrices of dimension $32{,}000$, whose full diagonalization reproduces the ground-state energy of the pentamer to good accuracy. The method is largely independent of the particular matrices used; how the matrix elements of this five-body system are computed is described in earlier work~\cite{base,AoyamaGEM2026}. On these precomputed matrices the BVM reproduces, as shown below, the reference energy of the full diagonalization to within $0.01$\,K with only $705$ basis functions and to within $0.001$\,K with $2{,}127$---a reduction of the matrix memory by $99.95\%$ and $99.56\%$, respectively. This reduction of the basis opens the way to realistic variational calculations for few-body systems with large degrees of freedom---six-body, seven-body, and beyond---that have been impractical with conventional approaches.

The remainder of this paper is organized as follows. Section~\ref{sec:method} describes the method: the discretized basis functions, the Gaussian-process surrogate model together with its acquisition function, the basis-selection algorithm, the incremental diagonalization, and the trimming procedure. Section~\ref{sec:validation} validates the GPR-based selection against purely random search in controlled single-step tests, including the sensitivity to the hyperparameters, and verifies the incremental diagonalization numerically. Section~\ref{sec:results} presents the full-scale results for the $^4$He pentamer. Section~\ref{sec:summary} gives a summary and conclusions.

\section{Method}
\label{sec:method}

In general, the space of candidate basis functions is infinite. In the GEM~\cite{Hiya2003} adopted in this work, each basis function is characterized by Gaussian width parameters of the relative motions, which are continuous variables. In addition, the so-called channels---the combinations of quantum numbers such as orbital angular momenta, spins, and isospins---can in principle be combined in infinitely many ways.

In practice, this space can be restricted by the physical properties of the system under consideration, such as its symmetries and spatial extent. Even so, the resulting candidate set remains large: for the five-body system treated in this paper, a realistic discretized candidate set typically exceeds $N_{\rm base} \sim 10^{5}$ functions, and the selection of a near-optimal subset of $n$ basis functions---say $n \sim 5000$---is a combinatorial optimization over ${}_{N_{\rm base}}C_{n}$ configurations, e.g.\ ${}_{100000}C_{5000} \approx 10^{8600}$ (Sec.~\ref{sec:intro}).

To make the problem tractable, we introduce the two concepts of a \emph{pool} and a \emph{tuple}, and replace the single intractable search by a sequence of small, feasible ones. The basis is built up iteratively, $K$ functions at a time, so that the accepted basis grows as $K \to 2K \to 3K \to \cdots \to n_{\rm step} K = n$. At each step, a pool of $N_{\rm pool}$ candidates ($N_{\rm pool} \ll N_{\rm base}$) is generated at random from the full candidate set. From this pool, we search for the $K$-element tuple that, when added to the basis accepted and fixed in the preceding steps, lowers the variational energy the most, by solving the generalized eigenvalue problem $Hc=ESc$ for the enlarged basis. In this way the full ${}_{N_{\rm base}}C_{n}$ optimization is reduced to a sequence of ${}_{N_{\rm pool}}C_{K}$ problems (e.g.\ ${}_{100}C_{5} \approx 7.5\times10^{7}$), each of which is small enough to be treated---though still far too large for exhaustive evaluation, which is the role of the Gaussian-process surrogate model introduced below. In actual applications, $n$ is usually not fixed in advance but is determined by monitoring the convergence of the energy.

Two operations dominate the cost of this procedure. First, the variational energy has to be evaluated for a large number of candidate tuples, and solving the full generalized eigenvalue problem from scratch for each candidate would require a prohibitive computational cost. We avoid this through an incremental diagonalization that reuses a stored factorization of the already accepted basis, so that each candidate is evaluated at $O(N^2)$ cost instead of by a full rediagonalization. Second, the number of possible candidate tuples is combinatorially large, so the strategy used to choose which candidates to evaluate has a strong impact on the overall efficiency. Rather than drawing candidates at random, we introduce a GPR surrogate model that predicts candidate energies and, combined with an acquisition function, preferentially evaluates the most promising tuples.

\subsection{Basis functions and matrix elements}
\label{subsec:basis}

In the present calculations we use, as a typical example, the basis of
the GEM~\cite{Hiya2003}, applied to the
$^4$He pentamer (a five-body system). The algorithm described in the
subsections that follow, however, is essentially independent of the
specific form of the basis functions and is general: what it actually
requires is only the elements of the Hamiltonian and overlap matrices,
$H$ and $S$, described by the variational parameters, and no
information on the particular basis functions is needed. Each basis function is a product of radial functions on the Jacobi coordinates multiplied by coupled spherical harmonics, the radial part being a Gaussian whose width parameters are taken on a geometric-progression grid,
\begin{equation}
\nu_j = 1/b_j^2,\qquad b_j = b_{\min}\,\gamma^{\,j-1}\quad(j=1,\dots,n_{\max}).
\end{equation}
Whereas the standard stochastic variational method treats these nonlinear parameters as continuous variables, restricting them to a fixed grid turns the basis into a finite set of candidate functions. This restriction also facilitates the direct comparison with exact reference solutions. The full candidate set used here contains $N_{\rm base}=32{,}000$ functions; its detailed construction---the channels, angular-momentum couplings, the $^4$He--$^4$He interaction, and the resulting energy convergence---is given in Ref.~\cite{base}.

For a chosen set of $N$ basis functions, the coefficients $c_j$ and the energy $E$ follow from the generalized eigenvalue problem $\sum_j (H_{ij}-E\,S_{ij})c_j=0$, where $H_{ij}$ and $S_{ij}$ are the Hamiltonian and overlap matrix elements of the basis functions. To save computational resources, the full $N_{\rm base}\times N_{\rm base}$ Hamiltonian and overlap matrices are computed once, stored, and loaded into memory, and are reused as a fixed, common input for all runs reported here. This not only avoids recomputing identical matrix elements across the many comparison experiments (different pool sizes $N_{\rm pool}$ and tuple sizes $K$, as well as trimming studies), but is in fact required for the speed comparisons to be well defined, since a meaningful speed comparison presupposes an identical input matrix. The present method then selects a small, near-optimal subset from this fixed candidate set. (In an actual application, the matrix elements would instead be evaluated successively as the calculation proceeds.)

Because the GEM basis is strongly non-orthogonal, with large mutual overlaps between functions, the overlap matrix of a selected subset is generally close to singular, which is handled by a test on the bordered overlap matrix---equivalent to the conventional canonical orthogonalization---introduced below. Following standard practice in canonical orthogonalization, the basis functions are first rescaled to unit diagonal overlap ($S_{ii}=1$); this leaves the eigenvalues unchanged while making the near-null directions easier to read off and keeping the matrix elements within a narrow range, which reduces rounding error. In addition, as a general safeguard against residual linear
dependence, any candidate evaluation that fails numerically or yields an
unphysically deep energy below a fixed floor $E_{\rm floor}$ is
discarded, as is any adopted tuple whose energy gain over the current
step departs markedly from the prevailing trend;
moreover, as the optimization proceeds and the accepted basis grows, a
function adopted at an early stage may gradually become expressible almost
completely as a superposition of functions adopted later; such a function
then spans no new variational space and is no longer needed. Redundancy of
this kind accumulates steadily and is removed most effectively by the
trimming procedure introduced below (Secs.~\ref{subsec:algorithm} and
\ref{subsec:trim}). We note that, within the scope of this paper, the
test equivalent to the canonical orthogonalization and the trimming
sufficed to remove the linear dependence.

\subsection{Gaussian-process surrogate model}
\label{subsec:gpr}
A GPR model was employed
to construct a surrogate model for the variational energy
associated with a candidate basis-function tuple.
Gaussian processes may be viewed as probabilistic
interpolation schemes \cite{Rasmussen2006}.
Instead of constructing a deterministic function that
exactly reproduces the teacher data, they assume that
the target values are generated from a correlated random
process. The correlation between two input points is
specified through a kernel function. Once the kernel is
defined, the values at unevaluated points can be inferred
from previously observed data together with an estimate
of the associated uncertainty. In the present work,
the Gaussian-process model is used to predict the
variational energies of unevaluated basis-function tuples. In Bayesian
terms, the kernel specifies the prior over functions, the evaluated
tuples constitute the data, entering through a noise-free likelihood
since their energies are deterministic, and the predictive mean and variance
introduced below are the posterior mean and the posterior uncertainty.

Let
\begin{equation}
A=\{i_1,i_2,\ldots,i_K\}
\end{equation}
denote a tuple consisting of $K$ candidate basis
functions selected from the current candidate pool.
The corresponding variational energy is regarded as a
realization of an unknown function,
\begin{equation}
E=f(A).
\end{equation}
The objective of the GPR model is to infer this function
from a finite set of previously evaluated tuples.

Unlike conventional regression problems, the input is not a
continuous vector but a discrete, unordered set of
basis-function indices. To apply GPR, each tuple must be
mapped onto a fixed-length numerical representation, and this
representation should not depend on the order in which the
indices are listed.

For this purpose we use a multi-hot representation: each
tuple is encoded as a binary vector whose length equals the
number $N_{\rm pool}$ of basis functions in the candidate
pool, with entries equal to one at the positions of the
selected basis functions and zero elsewhere. We denote the
multi-hot vector of a tuple $A$ by
$\bm{x}_A\in\{0,1\}^{N_{\rm pool}}$. For example, for a pool
of ten basis functions, the tuple consisting of the first,
third, and fifth functions is represented by the vector
$(1,0,1,0,1,0,0,0,0,0)$, while the tuple consisting of the
first, second, and fifth functions is represented by
$(1,1,0,0,1,0,0,0,0,0)$. By construction, this encoding is
invariant under reordering of the indices.

A natural measure of the similarity between two tuples $A$
and $B$ is the number of basis functions they have in common
---that is, the number of positions at which both multi-hot
vectors are equal to one. This overlap is the inner
product of the two vectors,
\begin{equation}
n_{\rm ov}(A,B)=\sum_{j=1}^{N_{\rm pool}}
x_{A,j}\,x_{B,j}
=\bm{x}_A^{T}\bm{x}_B
=|A\cap B| .
\end{equation}
In the example above, the two tuples share the first and
fifth basis functions, so $n_{\rm ov}=2$. This set-overlap
measure, rather than the Euclidean distance between the
vectors, is used to define the kernel introduced below.

Gaussian (radial basis function) kernels are the standard
choice in Gaussian-process regression \cite{Rasmussen2006}
and could in principle be applied here through the Euclidean
distance between the multi-hot vectors. For tuples of fixed
size $K$, however, this squared distance is determined
entirely by the overlap of the two tuples---for two tuples
sharing $c$ of their $K$ basis functions it equals
$2(K-c)$. The corresponding Gaussian kernel nevertheless
introduces an additional length-scale hyperparameter that
must be tuned, even though the underlying notion of
similarity is already completely specified by the tuple
overlap. We instead adopt the Tanimoto (Jaccard) kernel
\cite{Tanimoto1958,Bajusz2015}, which quantifies similarity
directly as the fraction of basis functions shared by two
tuples. It is a positive-definite kernel, is naturally
bounded between zero and one, and provides a similarity
measure tailored to set-valued inputs such as the present
basis-function tuples without introducing an additional
distance-scale parameter. As practical guidance for similar applications, we note
that a Gaussian kernel is not unusable here, but it parametrizes the
$K+1$ possible similarity values by an unnecessary extra degree of
freedom, so that a redundant hyperparameter search becomes unavoidable.
The Tanimoto kernel encodes the same information---the tuple
overlap---with no adjustable parameter at all; the insensitivity of the
results to the remaining nominal parameters is demonstrated in
Sec.~\ref{sec:val_hyper}. For set-valued inputs of this kind, the
Tanimoto kernel is often better suited than the more commonly used
Gaussian kernel.

For two tuples \(A\) and \(B\), the kernel function is defined as
\begin{equation}
k(A,B)
=
\sigma_f^2
\frac{|A\cap B|}
{|A\cup B|},
\end{equation}
where \(|A\cap B|\) and \(|A\cup B|\) denote the numbers of
elements in the intersection and union of the two tuples,
respectively, and \(\sigma_f\) is a scale parameter controlling
the overall covariance. In the actual calculations we set
$\sigma_f=1$: the teacher energies are standardized to zero mean
and unit variance before the regression, so the overall energy
scale is absorbed by the standardization and no tuning of
$\sigma_f$ is required.

In the present application, all tuples contain the same number
\(K\) of basis functions. The kernel therefore reduces to
\begin{equation}
k(A,B)
=
\sigma_f^2
\frac{n_{\rm ov}}
{2K-n_{\rm ov}},
\end{equation}
where $n_{\rm ov}$ is the overlap of the two tuples defined above.
The covariance is therefore completely determined by the number
of shared basis functions and increases monotonically with
$n_{\rm ov}$, reaching its maximum value when the two tuples are
identical.

Suppose that $n_t$ teacher tuples
\begin{equation}
\{A_i,E_i\},
\qquad
i=1,\ldots,n_t,
\end{equation}
have already been evaluated. The covariance matrix is given
by
\begin{equation}
M_{ij}
=
k(A_i,A_j)
+
\sigma_n^2\delta_{ij},
\end{equation}
where $\sigma_n$ is a small regularization parameter and
$\delta_{ij}$ denotes the Kronecker delta. We fix $\sigma_n^2=10^{-6}$, a value customarily used for Gaussian
processes of unit scale. Its role here is purely numerical---the
teacher energies are deterministic, converged eigenvalues, not noisy
observations---and any
value lying well above the machine-precision scale and well below the
smallest eigenvalue of the covariance matrix $M$ leaves the results
strictly unchanged, as verified in Sec.~\ref{sec:val_hyper}; in the
present setting this smallest eigenvalue is about $0.2$, so the
admissible window spans many orders of magnitude. Since $\sigma_f$ is likewise fixed to unity by
the standardization, as noted above, the Tanimoto-kernel model in the
present setting is effectively left with a single hyperparameter: the
exploration weight $\beta$ of the selection criterion introduced below.

The target energies are standardized according to
\begin{equation}
y_i
=
\frac{E_i-\bar E}{s_E},
\end{equation}
where \(\bar E\) and \(s_E\) denote the sample mean and standard deviation of the teacher energies,
\begin{equation}
\bar E=\frac{1}{n_t}\sum_{i=1}^{n_t}E_i,
\qquad
s_E=\sqrt{\frac{1}{n_t}\sum_{i=1}^{n_t}\left(E_i-\bar E\right)^{2}} .
\end{equation}

For an unevaluated tuple $A_*$, the covariance vector is defined as
\begin{equation}
\mathbf{k}_*
=
\left(
k(A_1,A_*),
k(A_2,A_*),
\ldots,
k(A_{n_t},A_*)
\right)^T .
\end{equation}
The predictive mean of the variational energy is then given by
\begin{equation}
\mu(A_*)
=
\bar E
+
s_E
\mathbf{k}_*^T
M^{-1}
\mathbf{y},
\end{equation}
while the predictive variance is
\begin{equation}
\sigma^2(A_*)
=
s_E^2
\left[
k(A_*,A_*)
-
\mathbf{k}_*^T
M^{-1}
\mathbf{k}_*
\right].
\end{equation}
The quantities $\mu(A_*)$ and $\sigma(A_*)$ represent
the predicted variational energy and its associated
uncertainty, respectively.

Because the goal is to find the tuples with the lowest variational
energy, the surrogate model is used to rank unevaluated tuples by
their lower confidence bound (LCB),
\begin{equation}
a(A_*) = \mu(A_*) - \beta\,\sigma(A_*),
\end{equation}
where $\beta \ge 0$ balances exploitation (selecting tuples with
low predicted energy $\mu$) against exploration (selecting tuples
with large predictive uncertainty $\sigma$). The quantity $a(A_*)$
may be interpreted as an optimistic estimate of the variational
energy, representing how low the energy might plausibly be. Tuples
with the smallest optimistic estimates are therefore regarded as
the most promising and are selected for explicit evaluation. The LCB is the acquisition function of Bayesian optimization;
it is this criterion that turns the GPR posterior into a decision rule for
which candidates to evaluate next. How strongly $\beta$ influences the
search also depends on how many of the top-ranked candidates are
evaluated in each round: when the evaluated group is large ($500$ in
this work), a moderate change of $\beta$ mostly reorders candidates
within the range that is evaluated anyway, so the outcome is expected
to be insensitive to $\beta$ (Sec.~\ref{sec:val_hyper}).

In each round, a large number of candidate tuples are scored by
the surrogate model. Evaluating the predictive variance for every
candidate requires solving the triangular linear system
$L\mathbf{x}=\mathbf{k}_*$ by substitution---an operation quadratic
rather than cubic in the teacher-set size, which we call a
\emph{triangular solve} throughout---where $M = LL^{T}$ is the Cholesky
factor of the covariance matrix, and this dominates the cost of the
acquisition step. By contrast, evaluating the predictive mean
requires only a single inner product of negligible cost,
$\mathbf{k}_*^{T}(M^{-1}\mathbf{y})$ with the precomputed vector
$M^{-1}\mathbf{y}$. To exploit this asymmetry, the acquisition
procedure is carried out in two stages.

Since the kernel satisfies $k(A_*,A_*)=\sigma_f^2$, the predictive
variance obeys
$0\le\sigma^2(A_*)\le s_E^2\sigma_f^2$, so that
$\sigma(A_*)\le s_E\sigma_f$. This gives a rigorous lower bound on
the optimistic estimate,
\begin{equation}
a(A_*)\ge a_{\rm low}(A_*)
=\mu(A_*)-\beta\,s_E\sigma_f .
\end{equation}

In the first stage, only the predictive mean is computed for all
candidates, and the $n_{\rm short}$ tuples with the smallest
$a_{\rm low}$ are retained as a short list. Since
$a_{\rm low}$ is a monotonically increasing function of $\mu$,
this amounts to selecting the candidates with the lowest predicted
energies. In the second stage, the exact predictive variance, and
hence the exact optimistic estimate, is computed only for the
tuples in the short list. Provided that $n_{\rm short}$ is large
enough that the short list contains every candidate whose
$a_{\rm low}$ lies below the optimistic estimate of the truly best
tuples, the two-stage procedure reproduces the ranking obtained
from the full evaluation at a small fraction of the cost, as
confirmed numerically. A related implementation remark concerns the teacher
side: within a step the teacher set only grows, so the old block of the
covariance matrix $M$ never changes. When a group of newly evaluated
tuples is appended (added together rather than one at a time), the
stored Cholesky factor is therefore not
recomputed but extended by the classical bordered update---one
triangular solve, one symmetric rank update, and one small factorization
of the new block---at a cost quadratic rather than cubic in the teacher
set size, and only the kernel entries of the new rows are built. Should
the small factorization fail owing to numerical linear dependence, the
factor is rebuilt from scratch. The same block structure serves the
parallel implementation: only the new rows of the factor need to be
communicated after each round. This bordered update is known as the
sequential update~\cite{Rasmussen2006}.

\subsection{Basis-selection algorithm}
\label{subsec:algorithm}

The complete algorithm is organized as two nested loops, supplemented by
an occasional housekeeping step:
\begin{itemize}
\item an \emph{outer loop} (the steps), each step of which enlarges the
variational basis by one $K$-tuple, so that the basis dimension grows as
$K \to 2K \to 3K \to \cdots$ (for $K=5$:
$5 \to 10 \to 15 \to \cdots$);
\item an \emph{inner loop} (the GPR rounds) inside each step, which
selects the $K$-tuple to be added by the GPR-guided Bayesian
optimization of Sec.~\ref{subsec:gpr};
\item a \emph{trimming} step, performed every fixed number of outer
steps, which removes basis functions that have become redundant.
\end{itemize}

One outer step proceeds as follows.

\begin{enumerate}

\item \emph{Pool construction.} A temporary pool of $N_{\rm pool}$ basis
functions is drawn at random from the candidate set. Functions already
accepted into the variational basis are excluded: once accepted, a
function is removed from the candidate set and cannot be drawn again.

\item \emph{Initial teacher set.} Random $K$-tuples are generated from
the pool, and for each tuple the variational energy of the enlarged
basis---the accepted basis plus the tuple---is evaluated from the
generalized eigenvalue problem
\begin{equation}
Hc = ESc .
\end{equation}
These evaluated tuples form the initial teacher set of the GPR model.
In the implementation the evaluation is incremental: the Cholesky factor
of the accepted overlap matrix and the accepted ground state are carried
over from the previous step---they are updated, not recomputed from
scratch, when a tuple is adopted---and each trial tuple is evaluated
through a bordered factorization combined with an iterative eigensolver
that takes the stored ground state as its initial guess (a warm start,
as opposed to a cold start from a random vector carrying no information
about the solution; Sec.~\ref{subsec:increment}); no $O(N^3)$
rediagonalization occurs at any point. The details of this carried-over
update are given in Sec.~\ref{subsec:increment}, and its numerical
verification in Sec.~\ref{subsec:on2verify}. This reduction of the dominant
cost is what makes the GPR-based search practical. Within each evaluation,
linear dependence of the candidate functions on the accepted basis is
detected through the Schur complement of the candidate overlap block, and
numerically dependent candidates are discarded (Sec.~\ref{subsec:increment}).

\item \emph{Inner loop (GPR rounds).} The inner loop maintains a growing
set of explicitly evaluated tuples and uses it, through the GPR model, to
decide which tuples to evaluate next. One round consists of three
operations.

First, the model is updated: from all tuples evaluated so far in
this step---at the beginning, the random teacher set of item~2---the
covariance matrix is built with the Tanimoto kernel, the energies are
standardized, and the Cholesky factorization of the covariance matrix is
computed (Sec.~\ref{subsec:gpr}). The updated model returns, for any
unevaluated tuple, a predicted energy $\mu$ and an uncertainty $\sigma$.

Second, candidates are scored. A large search set of unevaluated
$K$-tuples is generated at random from the pool, excluding the tuples
already evaluated. Since the number of possibilities is ${}_{N_{\rm pool}}C_{K}$ (e.g.,
about $7.5\times10^{7}$ for $5$-tuples from a pool of $100$), the search
set must be commensurately large; its size, however, is a trade-off
against computation time. In the sequential test of
Sec.~\ref{sec:val_main} the search set contains $10^{6}$ tuples per
round---far more than could ever be diagonalized explicitly, which is
precisely why a low-cost surrogate score is needed. For every member of the search set the
predicted energy $\mu$ is computed; this requires only a single inner
product per tuple. The uncertainty $\sigma$, whose computational cost
per tuple is far higher (a triangular solve), is computed only for a short list
preselected by $\mu$, following the two-stage screening of
Sec.~\ref{subsec:gpr}. The short-listed candidates are then ranked by
the optimistic estimate $a=\mu-\beta\sigma$: a tuple is promising either
because its predicted energy is low ($\mu$) or because its energy is
still highly uncertain ($\sigma$), and $\beta$ weighs the two.

Third, the most promising candidates---the group of tuples with the
lowest $a$---are diagonalized explicitly, again through the incremental
scheme of item~2, and the resulting energies are appended to the teacher
set; the covariance matrix and its Cholesky factor are updated
incrementally rather than rebuilt (Sec.~\ref{subsec:gpr}).

This propose--evaluate--update round is repeated a prescribed number of
times. With each round the model learns more about the low-energy region
of the pool, and the proposals concentrate there while the
$\sigma$-term keeps probing tuples about which little is known. All
tuples evaluated during the step---the initial random ones and every
proposed group---remain collected, and it is from this collection that
the next item selects.

\item \emph{Adoption.} Once the inner loop has terminated, the step is
closed by a single decision: among all tuples evaluated explicitly in
this step, the one yielding the lowest variational energy consistent
with the safeguards of Sec.~\ref{subsec:basis} is adopted,
and its $K$ basis functions are appended to the variational basis.
Upon adoption, the stored Cholesky factor and the ground-state vector of
the accepted basis are updated by the incremental $O(N^2)$ procedure of
Sec.~\ref{subsec:increment}, and the next step starts from these updated
quantities.

\item \emph{Trimming (every fixed number of steps).} Each accepted basis
function is tentatively removed and the variational energy is
recomputed; if the resulting energy increase is smaller than a
prescribed threshold, the function is discarded permanently. This
removes accumulated nearly linearly dependent components and reduces the
basis dimension, thereby lowering the cost of subsequent incremental
diagonalizations. Naively, each tentative removal would require
rediagonalizing the reduced basis from scratch; Sec.~\ref{subsec:trim}
shows that this decision, too, can be made in $O(N^2)$. The rationale is that a nearly dependent basis
function spans almost no new variational space, so its removal raises
the energy only marginally---precisely the property that the energy
criterion detects. As the steps proceed, functions whose contribution
has been superseded by later additions become redundant in this sense,
and trimming removes them continually. This also serves the central goal
of the present method: the search for a basis that is as small as
possible is greatly facilitated by continually eliminating such
redundancy.

\end{enumerate}

The outer loop is terminated when a user-defined stopping criterion is
satisfied. Typical examples include the energy improvement becoming
smaller than a prescribed threshold, the maximum number of steps being
reached, or the dimension of the variational basis reaching a prescribed
maximum value.

\subsection{Incremental diagonalization using a bordered Cholesky factorization}
\label{subsec:increment}

The computational bottleneck of the basis-selection procedure is the repeated
evaluation of candidate tuples. Let $N$ denote the number of already accepted
basis functions and $K$ the number of basis functions contained in a candidate
tuple. A straightforward evaluation of a candidate tuple would require solving
the generalized eigenvalue problem

\begin{equation}
Hc = ESc
\end{equation}

for the enlarged $(N+K)$-dimensional basis. Since the Gaussian-process search
requires the evaluation of a large number of candidate tuples, repeatedly
solving the full generalized eigenvalue problem from scratch would require
a prohibitive computational cost.

The remedy adopted here is to carry, in addition to the stored matrix
elements of the accepted basis ($H_{\rm oo}$ and $S_{\rm oo}$ below),
only two derived quantities across the steps: the
Cholesky factorization of the overlap matrix of the accepted basis,

\begin{equation}
S_{\rm oo}=L_S^{\vphantom{T}}L_S^{T}
\end{equation}

($L_S$ lower triangular), and the ground-state coefficient vector $\bm{c}_0$
of the accepted basis. Both are maintained incrementally at $O(N^2)$ cost per
step, as described below. No eigendecomposition of the accepted basis is stored.

Let a candidate tuple consist of the $K$ basis functions
$\chi_1,\ldots,\chi_K$, and let $\phi_1,\ldots,\phi_N$ denote the
accepted basis functions. Throughout this subsection, the indices
$i,j=1,\ldots,N$ label the accepted functions and $r,s=1,\ldots,K$ label
the candidate functions. In the enlarged basis
$(\phi_1,\ldots,\phi_N,\chi_1,\ldots,\chi_K)$, the $(N+K)$-dimensional
overlap and Hamiltonian matrices have the block form

\begin{equation}
S^{(N+K)}=
\begin{pmatrix}
S_{\rm oo} & S_{\rm on}\\[2pt]
S_{\rm on}^{T} & S_{\rm nn}
\end{pmatrix},
\qquad
H^{(N+K)}=
\begin{pmatrix}
H_{\rm oo} & H_{\rm on}\\[2pt]
H_{\rm on}^{T} & H_{\rm nn}
\end{pmatrix}.
\end{equation}

Here the old--old blocks $S_{\rm oo},H_{\rm oo}\in\mathbb{R}^{N\times N}$,
with elements $\langle\phi_i|\phi_j\rangle$ and
$\langle\phi_i|H|\phi_j\rangle$, are by far the largest part; they are
already stored from the preceding steps, together with the factor $L_S$. New for each candidate tuple is
only the thin border: the old--new blocks
$S_{\rm on},H_{\rm on}\in\mathbb{R}^{N\times K}$,

\begin{equation}
(S_{\rm on})_{ir}=\langle\phi_i|\chi_r\rangle ,
\qquad
(H_{\rm on})_{ir}=\langle\phi_i|H|\chi_r\rangle ,
\end{equation}

and the new--new blocks $S_{\rm nn},H_{\rm nn}\in\mathbb{R}^{K\times K}$,

\begin{equation}
(S_{\rm nn})_{rs}=\langle\chi_r|\chi_s\rangle ,
\qquad
(H_{\rm nn})_{rs}=\langle\chi_r|H|\chi_s\rangle .
\end{equation}

\paragraph*{Bordered factorization.}
The enlarged overlap matrix admits the bordered Cholesky factorization

\begin{equation}
S^{(N+K)}=L^{\vphantom{T}}L^{T},
\qquad
L=
\begin{pmatrix}
L_S & 0\\[2pt]
Y^{T} & L_{22}
\end{pmatrix},
\label{eq:borderchol}
\end{equation}

with

\begin{equation}
Y=L_S^{-1}S_{\rm on},
\qquad
L_{22}^{\vphantom{T}}L_{22}^{T}=S_{\rm nn}-Y^{T}Y .
\label{eq:schur}
\end{equation}

The stored factor $L_S$ enters unchanged; only the thin border is new,
requiring one triangular solve $Y=L_S^{-1}S_{\rm on}$
[$O(N^2 K)$] and the factorization of the $K\times K$ Schur complement
$S_{\rm nn}-Y^{T}Y$. The Schur complement measures the part of the
candidate functions that lies outside the space already spanned by the
accepted basis. Candidate tuples that are numerically linearly dependent
on the accepted basis are therefore detected and discarded at this
point: a tuple is rejected if the factorization of the Schur complement
fails or if a squared diagonal element of $L_{22}$ falls below a
prescribed cutoff. This is the concrete realization, in the Cholesky
representation, of the linear-dependence safeguard announced in
Sec.~\ref{subsec:basis}.

\paragraph*{Warm-started iterative eigensolution.}
With the bordered factor $L$ at hand, substituting $S^{(N+K)}=LL^{T}$
and introducing the transformed coefficients $\bm{\eta}=L^{T}\bm{c}$
turns the generalized problem into a standard one,

\begin{equation}
\hat H\,\bm{\eta}=E\,\bm{\eta},
\qquad
\hat H=L^{-1}H^{(N+K)}L^{-T},
\end{equation}

whose lowest eigenvalue---the candidate energy---is the minimum of the
Rayleigh quotient

\begin{equation}
E=\min_{\bm{\eta}\neq 0}
\frac{\bm{\eta}^{T}\hat H\bm{\eta}}{\bm{\eta}^{T}\bm{\eta}} .
\end{equation}

$\hat H$ is never formed explicitly; it is only applied to vectors, each
application consisting of two triangular solves with $L$ and one product
with $H^{(N+K)}$, i.e.\ $O(N^2)$. The minimum is evaluated by a locally
optimal block preconditioned conjugate gradient (LOBPCG)
iteration~\cite{Knyazev2001} warm-started from the stored ground state:
the initial vector is

\begin{equation}
\bm{\eta}_0=
\begin{pmatrix}
L_S^{T}\bm{c}_0\\[2pt]
0
\end{pmatrix},
\end{equation}

i.e.\ the current ground state embedded in the enlarged space. Because
adding $K\ (\ll N)$ functions perturbs the ground state only weakly,
this starting vector is already close to the solution, and the iteration
count is observed to be small and essentially independent of
$N$---an empirical finding, verified numerically in
Sec.~\ref{subsec:on2verify}, rather than a general guarantee. As long
as $K$ and the iteration count remain effectively constant as $N$
grows, one candidate evaluation therefore costs $O(N^2)$, the iteration
count entering only as the prefactor. Candidate evaluations are mutually
independent---each worker node holds a replica of $L_S$ and
$\bm{c}_0$---and are therefore parallelized without communication.

\paragraph*{Update upon adoption.}
When the winning tuple is adopted, the same bordered update,
Eqs.~(\ref{eq:borderchol})--(\ref{eq:schur}), extends the stored factor
permanently, $L_S\leftarrow L$, at $O(N^2 K)$ cost, and the ground-state
vector $\bm{c}_0$ is refreshed on the enlarged basis by the same
warm-started iteration. The next step then starts from these updated
quantities. The $O(N^2)$ scaling of the complete cycle is verified
numerically in Sec.~\ref{subsec:on2verify}.

\paragraph*{Extension to several lowest states.}
Although the presentation above targets the ground state, nothing in the
scheme is tied to a single eigenpair, and applications often require the
lowest few eigenvalues. The bordered factorization,
Eqs.~(\ref{eq:borderchol})--(\ref{eq:schur}), concerns only the overlap
matrix and is unchanged. The iterative eigensolver is used in its
block form~\cite{Knyazev2001}: to obtain the lowest $p$ eigenvalues,
$p$ vectors $\bm{\eta}_1,\ldots,\bm{\eta}_p$ are iterated
simultaneously, the Rayleigh--Ritz step being performed in the (at most)
$3p$-dimensional subspace spanned by the current vectors, their
residuals, and the previous search directions; by the variational
(Hylleraas--Undheim/MacDonald) theorem the resulting Ritz values bound
the lowest $p$ exact eigenvalues from above. Each iteration costs $p$
applications of $\hat H$, i.e.\ $O(pN^2)$, so the overall scaling is
preserved for small $p$. Alternatively, the states can be computed one
at a time with the constrained iteration already used for trimming
(Sec.~\ref{subsec:trim}), keeping each converged eigenvector as an
orthogonality constraint for the next. The warm start generalizes
verbatim: the $p$ eigenvectors of the previous step, embedded as
$(L_S^{T}\bm{c}_k,\,0)^{T}$, serve as the initial block, and upon
adoption all $p$ vectors are refreshed together with $L_S$. Only the
storage of $p$ coefficient vectors is added to the quantities carried
across the steps.

\subsection{Trimming to control near-linear dependence}
\label{subsec:trim}

The trimming step of Sec.~\ref{subsec:algorithm} exists because
near-linear dependence accumulates in the accepted basis as the outer loop
proceeds: a function adopted early may gradually become expressible as a
superposition of functions adopted later (Sec.~\ref{subsec:basis}).
Removing such functions keeps the basis dimension---and with it the cost
of every subsequent operation---small, and directly serves the central
goal of the method, the search for a basis that is as small as possible.

The cost question is the same as for candidate evaluation. Once the
basis dimension has grown to $N$, asking by how much the energy would
rise upon the removal of a single function, and answering it by solving
the reduced eigenvalue problem from scratch, is an $O(N^3)$ operation,
repeated for every removal candidate. The same strategy as in
Sec.~\ref{subsec:increment}---reuse the stored Cholesky factorization and a
warm-started iteration---reduces each assessment to $O(N^2)$ here as
well, under the same proviso as there: each iteration costs $O(N^2)$,
and the assessment is $O(N^2)$ as long as the iteration count, which
enters as the prefactor, remains effectively independent of $N$.

The starting point is the generalized eigenvalue problem of the
$N$-dimensional accepted basis,
\begin{equation}
H_{\rm oo}^{\vphantom{T}}\bm{c}=E\,S_{\rm oo}^{\vphantom{T}}\bm{c},
\end{equation}
with $\bm{c}\in\mathbb{R}^{N}$ the coefficient vector. Let
$S_{\rm oo}=L_S^{\vphantom{T}}L_S^{T}$ be the stored Cholesky
factorization of the overlap matrix ($L_S$ lower triangular, maintained across the steps, Sec.~\ref{subsec:increment}). Substituting this factorization and introducing the
transformed coefficients
\begin{equation}
\bm{\eta}=L_S^{T}\bm{c},
\qquad
\bm{c}=L_S^{-T}\bm{\eta},
\end{equation}
turns the generalized problem into a standard one,
\begin{equation}
\hat H\bm{\eta}=E\bm{\eta},
\qquad
\hat H=L_S^{-1}H_{\rm oo}^{\vphantom{1}}L_S^{-T},
\end{equation}
whose ground-state energy is the minimum of the Rayleigh quotient,
\begin{equation}
E=\min_{\bm{\eta}\neq 0}
\frac{\bm{\eta}^{T}\hat H\bm{\eta}}{\bm{\eta}^{T}\bm{\eta}} .
\end{equation}
($\hat H$ is never formed explicitly; it is only applied to vectors,
each application consisting of two triangular solves with $L_S$ and one
product with $H_{\rm oo}$, i.e.\ $O(N^2)$.)

The central observation is that removing the basis function $\phi_i$
is exactly one linear constraint in these coordinates. Removal means
restricting the variational space to coefficient vectors whose $i$-th
component vanishes, and this component can be written as
\begin{equation}
c_i=\bm{e}_i^{T}\bm{c}
=\bm{e}_i^{T}L_S^{-T}\bm{\eta}
=(L_S^{-1}\bm{e}_i)^{T}\bm{\eta}
=\bm{v}_i^{T}\bm{\eta},
\end{equation}
with $\bm{e}_i$ the $i$-th unit vector and $\bm{v}_i=L_S^{-1}\bm{e}_i$. The condition $c_i=0$ is
therefore the orthogonality constraint $\bm{\eta}\perp\bm{v}_i$, and the
ground-state energy of the basis without $\phi_i$ is the constrained
minimum
\begin{equation}
E_{-i}=\min_{\bm{\eta}\perp\bm{v}_i}
\frac{\bm{\eta}^{T}\hat H\bm{\eta}}{\bm{\eta}^{T}\bm{\eta}} .
\label{eq:trimmin}
\end{equation}
By the variational principle $E_{-i}\ge E$, so the energy rise
$E_{-i}-E$ is nonnegative.
A constrained LOBPCG iteration~\cite{Knyazev2001}, with the
$\bm{v}_j$ as constraint vectors, evaluates this
constrained minimum at a cost of $O(N^2)$ per iteration: the constraint vectors
are kept orthonormalized, the warm start is the current ground state
projected onto the constraint-compatible subspace, and the residual is
projected likewise in every iteration, so that the search never leaves
that subspace. The construction extends directly to several removals: with
$R$ the set of functions already marked for removal, the ground-state
energy of the basis without them is
\begin{equation}
E_{R}=\min_{\bm{\eta}\perp\bm{v}_j,\ j\in R}
\frac{\bm{\eta}^{T}\hat H\bm{\eta}}{\bm{\eta}^{T}\bm{\eta}} ,
\label{eq:trimminR}
\end{equation}
evaluated with one constraint vector per member of $R$;
Eq.~(\ref{eq:trimmin}) is the case $R=\{i\}$.

Examining all $N$ accepted functions would still be wasteful. As a
computational heuristic, the candidates are therefore preselected by
ranking the ground-state amplitudes $|c_{0,i}|$, where $\bm{c}_0$ is the
current ground-state coefficient vector, and only the smallest few are
examined (in the implementation, about three times the number of
removals allowed). A small amplitude suggests that removing the
corresponding function has little effect on the current ground state;
it is not by itself a measure of linear dependence---in a
non-orthogonal basis a nearly dependent function can carry a large,
cancelling amplitude. The amplitudes are compared on the unit-diagonal
normalization $S_{ii}=1$ of Sec.~\ref{subsec:basis}, which removes the
arbitrariness of the basis-function scaling; and the preselection never
decides a removal by itself, since the decision is always made by the
energy rise obtained from the constrained eigenvalue calculation, a
redundant function missed at one trimming event remaining a candidate
at later ones. The candidates in this short list are examined in order of increasing
amplitude. For each
candidate $i\notin R$ the constrained minimum $E_{R\cup\{i\}}$ is
evaluated under the constraints kept so far plus the candidate's own,
and the criterion is the incremental energy rise: if
$E_{R\cup\{i\}}-E_{R}$ falls below the trimming threshold
$\varepsilon$, the candidate is marked for removal,
$R\to R\cup\{i\}$, its constraint is kept, and the warm start is
updated (the constraint of a rejected candidate is discarded). Each
removal is thus limited individually; the cumulative change $E_R-E$ of
one trimming event is not itself thresholded, but since the constrained
subspaces are nested, the increments are nonnegative and the cumulative
change stays below the number of removals times $\varepsilon$. The number of removals per
trimming event is capped at a prescribed fraction of the number of
functions added since the previous event.

Once the removals are decided, the stored quantities are updated rather
than rebuilt: the corresponding rows and columns are deleted from
$H_{\rm oo}$ and $S_{\rm oo}$, the factor $L_S$ is downdated by an
$O(N^2)$ procedure per removed function, and the ground-state vector is
refreshed by a warm-started $O(N^2)$ iteration. The accuracy of the downdated factor
is verified by one additional $O(N^2)$ residual check per trimming
event:
with the probe vector $\bm{z}$ taken as the current ground-state
coefficient vector, the relative residual
$\|(L_S^{\vphantom{T}}L_S^{T}-S_{\rm oo})\bm{z}\|_2/
(1+\|S_{\rm oo}\bm{z}\|_2)$ is monitored against a fixed tolerance. No dense $O(N^3)$ eigensolution is
therefore required at any point of the trimming step. Moreover, the
number of examined candidates and the number of removals are bounded by
the caps introduced above---a fixed multiple of the functions added
since the previous trimming event, which is independent of $N$ for a
fixed trimming interval---so that, with the iteration count bounded as
before, the total cost of one trimming event also scales as $O(N^2)$.
The numerical verification is presented in
Sec.~\ref{subsec:on2verify}.

\section{Validation}
\label{sec:validation}
\begin{table*}[!t]
\centering
\caption{GPR-guided versus purely random search, grown from $500$ teacher
tuples to $2500$ over four propose--evaluate--update iterations
($N_{\rm pool}=100$, $K=5$, ten trials). The win rows quote GPR
wins--losses--ties. All energies are in K.}
\label{tab:val600}
\begin{tabular}{crrrcrrc}
\hline
trial & $E^{\rm teach}_{\min}$ (K) & $E^{\rm GPR}_{\min}$ (K) & $E^{\rm RND}_{\min}$ (K) & win &
$E^{\rm GPR}_{\rm mean}$ (K) & $E^{\rm RND}_{\rm mean}$ (K) & win \\
\hline
1  &  2.361 &  2.084 &  2.361 & GPR &  31.8 & 394.6 & GPR \\
2  & 18.749 &  9.813 & 12.797 & GPR &  40.1 & 276.0 & GPR \\
3  &  0.761 &  0.472 &  0.633 & GPR &  13.2 & 276.5 & GPR \\
4  &  2.042 &  1.281 &  1.939 & GPR &  19.5 & 271.3 & GPR \\
5  & 11.813 &  9.432 & 11.813 & GPR &  44.3 & 290.8 & GPR \\
6  &  6.006 &  3.426 &  5.809 & GPR &  19.0 & 332.0 & GPR \\
7  &  2.088 &  1.124 &  1.750 & GPR &  11.1 & 191.3 & GPR \\
8  &  6.566 &  4.853 &  5.448 & GPR &  24.8 & 325.1 & GPR \\
9  & 14.634 &  6.347 &  6.738 & GPR &  25.1 & 227.8 & GPR \\
10 &  3.496 &  2.815 &  2.813 & RND &  29.2 & 295.3 & GPR \\
\hline
mean &  6.851 &  4.165 &  5.210 & & 25.8 & 288.1 & \\
win  &         &         &         & 9--1--0 &        &          & 10--0--0 \\
\hline
\end{tabular}
\end{table*}
\begin{table}[htbp]
\centering
\caption{Ten-trial-mean progression of the GPR-guided and purely random
searches from $n_t=500$ to $2500$ evaluated tuples ($N_{\rm pool}=100$, $K=5$): the
cumulative lowest energy $E_{\min}$ and the per-round candidate mean
$E_{\rm mean}$ for each side. All energies are in K.}
\label{tab:valprog}
\begin{tabular}{lrrrrr}
\hline
stage & $n_t$ & $E^{\rm GPR}_{\min}$ & $E^{\rm RND}_{\min}$ &
$E^{\rm GPR}_{\rm mean}$ & $E^{\rm RND}_{\rm mean}$ \\
 & & (K) & (K) & (K) & (K) \\
\hline
teacher & 500  & 6.851 & 6.851 & --- & --- \\
round 1 & 1000 & 4.278 & 5.934 & 27.4 & 297.6 \\
round 2 & 1500 & 4.240 & 5.846 & 46.4 & 305.8 \\
round 3 & 2000 & 4.223 & 5.225 & 26.6 & 295.8 \\
round 4 & 2500 & 4.165 & 5.210 & 25.8 & 288.1 \\
\hline
\end{tabular}
\end{table}

\begin{table*}[htbp]
\centering
\caption{Dependence on the tuple size $K$: the sequential comparison of
Table~\ref{tab:val600} (teacher $500\to2500$, search set $10^{6}$, ten
trials) repeated for $K=3,5,7,9,11,13,15$. Ten-trial means of $E_{\min}$
(cumulative lowest energy after $2500$ evaluated tuples) and
$E_{\rm mean}$ (mean energy of the $500$ tuples of the last round), the
GPR wins--losses--ties in $E_{\min}$, and the search-set coverage
$f_{\rm s}$ (see text). All energies are in K.}
\label{tab:valK}
\begin{tabular}{crrrcrr}
\hline
 & & \multicolumn{2}{c}{$E_{\min}$ (K)} & &
\multicolumn{2}{c}{$E_{\rm mean}$ (K)} \\
\cline{3-4}\cline{6-7}
$K$ & $f_{\rm s}$ (\%) & $E^{\rm GPR}_{\min}$ & $E^{\rm RND}_{\min}$ &
wins & $E^{\rm GPR}_{\rm mean}$ & $E^{\rm RND}_{\rm mean}$ \\
\hline
3  & $100$ ($6.2\times10^{2}$) & 4.758 & 7.005 & 9--1--0 & 47.0 & 571.8 \\
5  & $1.3$                & 4.165 & 5.210 & 9--1--0 & 25.8 & 288.1 \\
7  & $6.2\times10^{-3}$   & 3.561 & 4.883 & 10--0--0 & 18.4 & 189.6 \\
9  & $5.3\times10^{-5}$   & 3.422 & 4.238 & 10--0--0 & 13.8 & 132.4 \\
11 & $7.1\times10^{-7}$   & 3.247 & 3.982 & 9--0--1 & 11.6 & 100.8 \\
13 & $1.4\times10^{-8}$   & 3.175 & 3.725 & 10--0--0 & 9.8 & 79.8 \\
15 & $3.9\times10^{-10}$  & 3.067 & 3.429 & 9--1--0 & 8.7 & 65.3 \\
\hline
\end{tabular}
\end{table*}

Before turning to the full-scale calculations, we validate the two central
ingredients of the BVM introduced in Sec.~\ref{sec:method}: the GPR-guided selection of
candidate tuples and the incremental diagonalization. The
GPR-guided selection is validated in the subsections below by isolating
single steps of the algorithm and comparing it, over a set of independent
trials, with a purely random search that performs the same number of
evaluations without the surrogate. Throughout, the tests use the fixed $H$ and $S$
matrices (Sec.~\ref{subsec:basis}) and, except for the sensitivity study
of Sec.~\ref{sec:val_hyper}, the same fixed hyperparameters
($\sigma_f=1$, $\sigma_n^2=10^{-6}$, $\beta=2.0$; Sec.~\ref{subsec:gpr}).
Because the two searches are compared at an equal number of explicit
evaluations, the comparison isolates the quality of the selection itself,
independently of the implementation and its timing. The incremental
diagonalization is validated separately: its accuracy and its $O(N^2)$ cost
are verified in Sec.~\ref{subsec:on2verify}. All tests in this section are
carried out with a Python implementation of the method, which is well
suited to the numerical libraries and the data handling that the
validation requires and allows easy code development; the calculations of
Sec.~\ref{sec:results} run in a supercomputer environment and therefore
use an algorithmically equivalent Fortran implementation that reproduces
the results of the Python code.

\subsection{GPR-guided versus random selection}
\label{sec:val_main}

In this section we verify that, when selecting a tuple of $K$ basis
functions from a pool of $N_{\rm pool}$ candidate functions, a GPR-guided
selection that learns which tuples lower the energy is more efficient than
choosing them at random. As typical values we take $N_{\rm pool}=100$ and
$K=5$.

The comparisons of this and the following two subsections share a common
protocol. In each trial the pool and the sampled tuples are redrawn, and
the two searches---the GPR-guided one and the purely random one---each
return the lowest energy they reach. For each comparison we report, over
the ten independent trials, the number in which the GPR-guided search
reaches the lower energy. In the tables these counts are quoted as
wins--losses--ties for the GPR side; a tie means that the two sides end at
exactly the same lowest energy (in every tie observed in this work, neither
side improved on the minimum of the common teacher set).

Starting from a common initial teacher set of $500$ randomly generated
$K$-tuples of basis functions ($N_{\rm teacher}=500$), the
propose--evaluate--update cycle is repeated four times, each time adding
$500$ new tuples, so that the cumulative number of evaluated tuples grows
$500\to1000\to1500\to2000\to2500$. At each iteration the GPR model, now
conditioned on all tuples evaluated so far, proposes $500$ new candidates.
The full candidate space contains ${}_{100}C_{5}\approx7.5\times10^{7}$
tuples; even with the low per-candidate cost of the GPR surrogate, an
exhaustive scan remains impractical, so the search set at each round is
restricted to $10^{6}$ tuples (about $1.3\%$ of the total). This proposal undergoes the two-stage
screening described in Sec.~\ref{subsec:gpr}, carried out in parallel across
$500$ concurrent search processes. The random side draws $500$ additional tuples per iteration
independently and uniformly at random from the same pool, with no
information from the GPR model. Both sides start from the same teacher
set of $N_{\rm teacher}$ tuples rather than two independently drawn sets of
$N_{\rm teacher}$, so that any difference between $E^{\rm GPR}_{\min}$ and
$E^{\rm RND}_{\min}$ below is due entirely to the $2000$ further tuples each
side adds afterward, not to a difference in the starting point.

Table~\ref{tab:val600} lists, for each of ten independent trials,
$E^{\rm teach}_{\min}$ (the lowest energy among the initial $N_{\rm teacher}$
teacher tuples), $E^{\rm GPR}_{\min}$ and $E^{\rm RND}_{\min}$ (the cumulative
lowest energy of each side after $2500$ evaluated tuples), and
$E^{\rm GPR}_{\rm mean}$ and $E^{\rm RND}_{\rm mean}$, the mean of the
$500$ individual ground-state energies produced by the last iteration alone
(evaluated tuples $2000\to2500$), which show the typical rather than best-case quality
of those $500$ candidates: $500$ purely random candidates sample the bulk of
the energy distribution, while the $500$ GPR-proposed candidates are
concentrated near the low-energy tail targeted by the search.

The GPR side reaches the lower cumulative minimum in $9$ of the $10$
trials (mean $4.16$\,K versus $5.21$\,K); in trial $10$ the random side ends
marginally lower ($2.8132$\,K versus $2.8147$\,K, a difference of
$0.0015$\,K). On the mean of the final $500$ candidates the GPR side wins
all $10$ trials (mean $25.8$\,K versus $288.1$\,K).

Table~\ref{tab:valprog} lists the ten-trial means of the two searches at
each stage from $500$ to $2500$ evaluated tuples, resolving the single row of
Table~\ref{tab:val600} into its four propose--evaluate--update rounds at
$N_{\rm pool}=100$ and $K=5$. For each stage it reports, for the
GPR-guided and the purely random side, the cumulative lowest energy reached
up to that stage, $E_{\min}$, and the mean energy of the $500$ candidates
added in that round, $E_{\rm mean}$. At the initial $500$ tuples the two sides share the
same teacher set, so their cumulative minima coincide at $6.85$\,K and no
round-specific mean is defined there. The cumulative minimum of the GPR side
drops sharply in the first round and then improves only slowly,
$6.85\to4.28\to4.24\to4.22\to4.16$\,K; the random side falls to $5.93$\,K in
the first round and creeps down to $5.21$\,K thereafter, never approaching
the GPR level. The per-round mean of the $500$ freshly added
candidates tells the same story in a complementary way: the GPR proposals
stay in the range $26$--$46$\,K at every round---roughly an order of magnitude
below the random draws ($\approx288$--$306$\,K)---so the concentration of the
GPR proposals in the low-energy tail is a property of every iteration, not
only of the final one reported in Table~\ref{tab:val600}.
\begin{table*}[htbp]
\centering
\caption{Dependence on the pool size $N_{\rm pool}$: the protocol of
Table~\ref{tab:valK} repeated at fixed $K=5$ for $N_{\rm pool}=50$ to
$5000$ (ten trials). Ten-trial means of the search fraction
$f_{\rm s}=10^{6}/{}_{N_{\rm pool}}C_{5}$, $E_{\min}$ (cumulative
lowest energy after $2500$ evaluated tuples), the GPR wins--losses--ties
in $E_{\min}$, and $E_{\rm mean}$ (mean energy of the $500$ candidates
of the final round). All energies are in K.}
\label{tab:valpool}
\begin{tabular}{rrrrcrr}
\hline
 & & \multicolumn{2}{c}{$E_{\min}$ (K)} & & \multicolumn{2}{c}{$E_{\rm mean}$ (K)} \\
\cline{3-4}\cline{6-7}
$N_{\rm pool}$ & $f_{\rm s}$ (\%) & $E^{\rm GPR}_{\min}$ & $E^{\rm RND}_{\min}$ &
wins & $E^{\rm GPR}_{\rm mean}$ & $E^{\rm RND}_{\rm mean}$ \\
\hline
50   & $47$ & 14.583 & 16.266 & 8--1--1 & 42.6 & 300.6 \\
100  & $1.3$ & 4.165 & 5.210 & 9--1--0 & 25.8 & 288.1 \\
200  & $3.9\times10^{-2}$ & 3.265 & 5.350 & 10--0--0 & 24.3 & 311.2 \\
500  & $3.9\times10^{-4}$ & 1.303 & 2.437 & 9--1--0 & 20.2 & 321.5 \\
1000 & $1.2\times10^{-5}$ & 0.737 & 1.679 & 10--0--0 & 17.8 & 313.1 \\
3000 & $5.0\times10^{-8}$ & 0.913 & 2.676 & 10--0--0 & 22.7 & 314.1 \\
5000 & $3.8\times10^{-9}$ & 0.907 & 2.225 & 9--0--1 & 25.7 & 317.0 \\
\hline
\end{tabular}
\end{table*}

\subsection{Dependence on the tuple size $K$}
\label{sec:val_K}

The tuple size $K$ is the number of basis functions proposed and evaluated
together in one outer iteration, and it enters the procedure through two
competing effects. On one hand, a larger $K$ reaches a given basis dimension
in fewer outer loops, so if the full candidate set
${}_{N_{\rm pool}}C_{K}$ could be enumerated exhaustively, a larger $K$ would
be the more efficient choice. In practice, even with the GPR surrogate
standing in for the direct evaluation, the search cannot cover an arbitrarily
large candidate set; balancing this coverage against the overall computation
time, the present calculations adopt a search set of $10^{6}$ tuples. Seen as a search problem for the GPR,
increasing $K$ makes ${}_{N_{\rm pool}}C_{K}$ grow rapidly, so a fixed search
set of $10^{6}$ tuples covers a progressively smaller fraction
$f_{\rm s}\equiv 10^{6}/{}_{N_{\rm pool}}C_{K}$ of the whole
space; the probability that this set contains the true optimum decreases
accordingly, and the fraction of proposed tuples carrying basis functions of
little variational value grows, which runs against the aim of reaching the
lowest energy with as few basis functions as possible. Counterbalancing this,
at a fixed $K$ the GPR always selects the lowest-predicted-energy tuples
within whatever search set it is given, so its proposals should reach a lower
energy than purely random sampling at the same $K$. Enlarging $K$ to add
basis functions in larger groups and letting the GPR locate the best tuple
at a fixed $K$ therefore stand in a trade-off, and this subsection follows
both $E_{\min}$ and $E_{\rm mean}$ as $K$ is varied to see how it plays out.

Table~\ref{tab:valK} repeats the sequential comparison of
Sec.~\ref{sec:val_main} (teacher $500\to2500$ over four
propose--evaluate--update rounds, search set $10^{6}$) for tuple sizes
$K=3,5,7,9,11,13,15$, with all other conditions unchanged (the same ten
pools; the $10^{6}$ search set regenerated for each $K$). For each $K$ the
table lists two quantities, each averaged over the ten trials: the cumulative
lowest energy $E_{\min}$ reached after $2500$ evaluated tuples, and the mean
energy $E_{\rm mean}$ of the $500$ tuples proposed in the last round
($2000\to2500$), each given for the GPR-guided ($E^{\rm GPR}$) and the random
($E^{\rm RND}$) selection. It also lists the number of trials in which GPR
reaches the lower $E_{\min}$, and the search fraction $f_{\rm s}$. Several observations can be made. First, the advantage of the
GPR-guided selection on $E_{\min}$ is statistically significant for every $K$
examined ($9$--$10$ wins). Second, the attained minimum
energy decreases with $K$, from $4.76$\,K at $K=3$ to $3.07$\,K at $K=15$,
consistent with the variational principle. Third, the group mean
$E_{\rm mean}$ also decreases with $K$ for both methods, and
$E^{\rm GPR}_{\rm mean}$ lies well below $E^{\rm RND}_{\rm mean}$ at every $K$
(for instance $25.8$\,K versus $288.1$\,K at $K=5$), so the GPR proposals are
concentrated on lower-energy tuples than random sampling throughout the run,
not only at the single best tuple. Fourth, the difference between
$E^{\rm RND}_{\min}$ and $E^{\rm GPR}_{\min}$ tends to narrow as $K$ grows:
the GPR-guided search ends $2.25$\,K below the random one at $K=3$, but only
$0.36$\,K below at $K=15$. This is consistent with the discussion above: as
$f_{\rm s}$ decreases, the fixed search set gives the GPR-guided selection
less room to improve on random sampling. We note that the fixed search set of
$10^{6}$ tuples is effectively exhaustive for $K=3$: only
${}_{100}C_{3}\approx1.6\times10^{5}$ distinct $3$-tuples exist, so the
nominal ratio $10^{6}/{}_{100}C_{3}$---quoted in parentheses in
Table~\ref{tab:valK}---exceeds unity while the effective coverage is
$100\,\%$. At the other end, $f_{\rm s}$ shrinks to
$\sim4\times10^{-10}\,\%$ at $K=15$.

\subsection{Dependence on the pool size $N_{\rm pool}$}
\label{sec:val_pool}

Having shown in Table~\ref{tab:valK} that the GPR-guided selection outperforms
random search across tuple sizes, we next ask how the advantage depends on the
pool size $N_{\rm pool}$, the number of candidate basis functions from which each
$K$-tuple is drawn. Two competing effects are expected. A larger pool contains
more---and on average better---candidate functions, so the lowest energy
attainable by a single $K=5$ tuple should decrease. At the same time, with the
search set fixed at $10^{6}$ tuples, a larger pool means that a smaller fraction
$f_{\rm s}$ of the combinatorial space ${}_{N_{\rm pool}}C_{5}$ is examined, so
the surrogate has a lower chance of locating the true minimum. We also
note the computational costs associated with the pool size. Since
$N_{\rm pool}$ is the length of the multi-hot vector that represents
each tuple, enlarging the pool increases the cost of the kernel
evaluations; as the accepted dimension $N$ grows, however, this
contribution to the overall cost becomes progressively less important.
In addition, in actual applications the Hamiltonian and overlap matrix
elements between the accepted basis functions and the freshly drawn
pool candidates are in many cases computed anew at every step; taking
the time for these matrix elements into account as well, a smaller
$N_{\rm pool}$ is often preferable with respect to the total
computational cost.
Here we examine which pool size is appropriate.

The GPR minimum $E^{\rm GPR}_{\min}$ decreases rapidly with pool size, from
$14.6$\,K at $N_{\rm pool}=50$ to $0.74$\,K at $N_{\rm pool}=1000$, as the
enlarged pool makes lower-energy tuples available. Beyond about $1000$ the
improvement saturates: the values at $N_{\rm pool}=3000$ and $5000$
($0.91$\,K for both) are marginally higher than at $1000$, but with ten trials
per point we do not regard this small difference as significant. The
saturation is naturally attributed to the shrinking search fraction: once the
pool is so large that the fixed $10^{6}$ search set covers only a negligible
part of ${}_{N_{\rm pool}}C_{5}$ ($f_{\rm s}\lesssim10^{-7}\,\%$), the richer
candidate set no longer translates into a lower reached energy. Exploiting a
still larger pool would therefore require a correspondingly larger search set
or teacher set.

Across the whole range the GPR side reaches the lower $E_{\min}$ in $8$--$10$
of the ten trials, so the surrogate-guided search keeps its advantage over
random selection up to the largest pool (one trial each at
$N_{\rm pool}=50$ and $5000$ ends in a tie, in which neither side improves on
the teacher minimum). The two candidate means give a complementary picture.
The random mean $E^{\rm RND}_{\rm mean}$ stays close to $300$\,K throughout, as
expected once the $500$ random draws sample the bulk of the energy
distribution, so that the mean becomes essentially independent of the pool
size. The GPR mean $E^{\rm GPR}_{\rm mean}$, by contrast, decreases from about
$43$\,K at $N_{\rm pool}=50$ to about $18$\,K at $1000$ and then rises mildly
(to about $26$\,K at $5000$), following the same saturation of the
search-fraction effect.

The appropriate pool size depends on the system at hand. In the present
setting the attainable $E_{\min}$ improves rapidly up to
$N_{\rm pool}\approx1000$ and shows no further benefit beyond that; since a
larger pool also lengthens the multi-hot representation that describes a
tuple and, in an actual application where the matrix elements are
computed as they are needed, increases the number of matrix elements to
be computed at every step, without lowering the
reached energy, there is no reason here to go beyond a pool of about one
thousand.

\subsection{Hyperparameter sensitivity}
\label{sec:val_hyper}

The GPR model with the Tanimoto kernel involves three hyperparameters: $\sigma_f$, $\sigma_n^2$,
and $\beta$. As discussed in Sec.~\ref{sec:method}, the first two can be fixed from the
outset. The standardization of the teacher energies absorbs the overall
energy scale, so that $\sigma_f=1$ requires no tuning; and $\sigma_n^2$
is a purely numerical regularization, for which any value well above the
machine-precision scale and well below the smallest eigenvalue of the
covariance matrix (about $0.2$ in the present setting) is equivalent, so
that the customary $\sigma_n^2=10^{-6}$ can be adopted. The only
hyperparameter that could genuinely influence the search in this problem
is the exploration weight $\beta$ of the selection criterion
$a=\mu-\beta\sigma$: for $\beta=0$ the candidates are ranked purely by
their predicted energies (pure exploitation), while increasing $\beta$
shifts the preference toward candidates whose energies are still
uncertain (exploration).

We therefore examined the effect of $\beta$ directly. For each
$\beta=0,1,2,3,4$ the complete search of Sec.~\ref{sec:val_main}---the
teacher set of $500$ tuples followed by four propose--evaluate--update
rounds of $500$ candidates each, up to the final total of $2500$
evaluated tuples---was
repeated from the start with that $\beta$, in each of the ten trials
($N_{\rm pool}=100$, $K=5$; identical pools, teacher sets, and random
sequences for all $\beta$). Table~\ref{tab:valbeta} lists the
cumulative lowest energy at the final total: the reached energy
depends only weakly on $\beta$.

\begin{table}[htbp]
\centering
\caption{Sensitivity to the exploration weight $\beta$: the sequential
search of Sec.~\ref{sec:val_main} ($N_{\rm pool}=100$, $K=5$, teacher
$500\to2500$, ten trials) repeated from the start with each $\beta$
under otherwise identical conditions (see text). Entries are the
cumulative lowest energies $E_{\min}$ after $2500$ evaluated tuples,
in K.}
\label{tab:valbeta}
\begin{tabular}{crrrrr}
\hline
trial & $\beta=0$ & $\beta=1$ & $\beta=2$ & $\beta=3$ & $\beta=4$ \\
\hline
1  & 1.985 & 2.084 & 2.084 & 2.084 & 2.084 \\
2  & 9.813 & 9.813 & 9.813 & 9.813 & 9.813 \\
3  & 0.472 & 0.472 & 0.472 & 0.485 & 0.485 \\
4  & 1.281 & 1.281 & 1.281 & 1.285 & 1.489 \\
5  & 9.023 & 9.432 & 9.432 & 9.432 & 9.432 \\
6  & 3.426 & 3.426 & 3.426 & 3.426 & 3.426 \\
7  & 1.124 & 1.124 & 1.124 & 1.124 & 1.124 \\
8  & 4.707 & 4.707 & 4.853 & 4.855 & 4.855 \\
9  & 5.730 & 6.347 & 6.347 & 6.305 & 6.282 \\
10 & 2.588 & 2.815 & 2.815 & 2.580 & 2.580 \\
\hline
mean & 4.015 & 4.150 & 4.165 & 4.139 & 4.157 \\
\hline
\end{tabular}
\end{table}
\begin{table}[!t]
\centering
\caption{Timing $t$ of the standard two-stage diagonalization as a
function of the basis dimension $N$, and the effective exponent $p$
between adjacent sizes (see text). Times are in seconds.}
\label{tab:on3base}
\begin{tabular}{rrr}
\hline
$N$ & $t$ [s] & $p$ \\
\hline
1000  &   0.17 & --  \\
2000  &   1.05 & 2.7 \\
3000  &   3.73 & 3.1 \\
4000  &   8.74 & 3.0 \\
5000  &  17.99 & 3.2 \\
10000 & 154.0  & 3.1 \\
\hline
\end{tabular}
\end{table}
\begin{table*}[htbp]
\centering
\caption{Incremental update versus standard diagonalization for four
adoption steps $N_1\to N_2$ ($K=N_2-N_1$ functions added): the total
incremental time $t_{\rm inc}$ (LOBPCG iteration count in parentheses),
the time $t_{\rm std}$ of the standard diagonalization of the $N_2$
basis, and the ground-state energy difference
$\Delta E=E_{\rm inc}-E_{\rm std}$. Times are in seconds; energies are
in K.}
\label{tab:on2verify}
\begin{tabular}{ccrrr}
\hline
$N_1\to N_2$ & $K$ & $t_{\rm inc}$ [s] (iterations) & $t_{\rm std}$ [s] &
$\Delta E$ (K) \\
\hline
$1500\to1600$   & 100  &  1.05 (575) &   0.56 & $+3\times10^{-12}$ \\
$3000\to3100$   & 100  &  3.80 (553) &   4.14 & $+2\times10^{-11}$ \\
$5000\to5100$   & 100  &  9.81 (544) &  19.58 & $+2\times10^{-11}$ \\
$10000\to11000$ & 1000 & 46.7 (565)  & 209.9  & $-2\times10^{-10}$ \\
\hline
\end{tabular}
\end{table*}

Two circumstances underlie this insensitivity. In the early rounds the
teacher tuples are sparse in the space of $7.5\times10^{7}$ candidates:
most candidates share at most one basis function with any teacher tuple
(kernel values $0$ or $1/9$), the predictive $\sigma$ is pinned near
its upper bound $s_E\sigma_f$, and the LCB ranking coincides with the
$\mu$ ranking regardless of $\beta$. In the later rounds, when teacher
data accumulate near the minimum and $\sigma$ does differentiate
between candidates, the protection comes from the size of the evaluated
group, as anticipated in Sec.~\ref{subsec:gpr}: a change of $\beta$
mostly reorders candidates within the top group of $500$ that is
evaluated anyway. This protection is lost when the evaluated group is
small: for example, with $100$ candidates evaluated per round, the
late-round energies vary appreciably with $\beta$.

On this basis we adopt $\sigma_f=1.0$, $\sigma_n^2=10^{-6}$, and
$\beta=2.0$---an intermediate value of the insensitive range, already
used in the preceding subsections---and keep them fixed in all
calculations that follow.

\subsection{Numerical verification of the $O(N^2)$ incremental update}
\label{subsec:on2verify}

The incremental update of Sec.~\ref{subsec:increment} consists of the
bordered update of the stored Cholesky factor $L_S$,
Eqs.~(\ref{eq:borderchol})--(\ref{eq:schur}), followed by the
warm-started LOBPCG iteration. Each of these operations costs $O(N^2)$
per iteration, so one adoption step costs $O(N^2)$ in total as long as
the LOBPCG iteration count does not grow with $N$. Under this premise
the iteration count enters the cost only as a constant prefactor
$\alpha$, and $\alpha N^{2}$ is $O(N^{2})$ however large $\alpha$ may
be. In this subsection the verification is carried out up to basis
dimensions of order $10^{4}$. The applications envisaged for the
method---six- and seven-body systems---involve dimensions of order
$N\sim10^{5}$--$10^{6}$, and we consider the present range sufficient
for establishing the scaling. At those dimensions the difference in
exponent is decisive: the ratio between the $O(N^{3})$ operation count
of the standard two-stage diagonalization and the $\alpha N^{2}$
operations of the incremental update, with $\alpha$ of a few hundred as
measured below, grows linearly with $N$. We note that the present work
reuses precomputed matrix elements to save computational cost
(Sec.~\ref{subsec:basis}); even if the matrix elements were instead
computed as they are needed, the border elements required in one step
cost only $O(N)$ to evaluate, so in either case the total computational
cost is dominated by the diagonalization.

This subsection verifies these claims numerically. The incremental
update is compared with the standard two-stage diagonalization (the
canonical orthogonalization of Sec.~\ref{subsec:basis}) as the
reference, and three points are examined: (i) that the incremental
update entails no loss of accuracy; (ii) that the wall-clock times scale with
the basis dimension $N$ as $O(N^{2})$ and $O(N^{3})$, respectively,
quantified by the effective exponent $p=\ln(t_2/t_1)/\ln(N_2/N_1)$
obtained from the measured times; and (iii) that the LOBPCG iteration
count does not depend on $N$. Both methods were timed in a
single-process implementation on the same computer (Apple M3 Pro
processor, 18~GB unified memory), under identical
conditions, each entry being the median of three runs; the meaningful
quantities are the effective exponents and the time ratio between the
two methods, not the absolute times, which depend on the machine (the
full-scale calculations of Sec.~\ref{sec:results} employ the same incremental scheme
in its parallel form).

The verification uses the precomputed matrices of the $32{,}000$
candidate functions of Sec.~\ref{subsec:basis}. From these, $12{,}000$
functions are drawn in a fixed random order and rescaled
to unit diagonal overlap as in Sec.~\ref{subsec:basis}. The functions are
then adopted in this order through the bordered update of
Sec.~\ref{subsec:increment}, subject to its Schur-complement test
[Eq.~(\ref{eq:schur}), cutoff $10^{-10}$], until the $11{,}000$ functions
required for the present verification are collected.

Table~\ref{tab:on3base} first establishes the reference scaling of the
standard diagonalization, performed in the two stages of canonical
orthogonalization: the overlap matrix is diagonalized, eigenvalues
$\le10^{-12}$ are discarded, and the Hamiltonian projected onto the
remaining directions is diagonalized. The effective exponent between
adjacent sizes lies in the range $p=2.7$--$3.2$, and the range
$N=2000$--$10000$ as a whole gives $p\approx3.1$, consistent with the
$O(N^{3})$ operation count.

The incremental update is then timed against this reference for four
steps $N_1\to N_2$: $1500\to1600$, $3000\to3100$, and $5000\to5100$
(each adding $K=100$ functions), and $10000\to11000$ ($K=1000$). In each
step the stored factor $L_S$ and ground-state vector $\bm{c}_0$ of the
$N_1$ basis are reused, the border of $K$ functions is added by the
bordered update and the warm-started LOBPCG iteration, and the result is
compared with a standard diagonalization of the same $N_2$ basis
(Table~\ref{tab:on2verify}). A border of $K=100$ rather than the
value $K=5$ of the full-scale calculations is used because at $K=5$ the timing difference is
buried in the run-to-run fluctuations of the library routines; the last
step probes a large border ($K=1000$) at the dimension of the envisaged
applications.

Four observations follow from Table~\ref{tab:on2verify}. (i) The two
methods agree: the ground-state energies differ by
$|\Delta E|\le2\times10^{-10}$\,K in every step, including the largest
one with $K=1000$, so the incremental update entails no loss of
accuracy. (ii) The scaling is as designed. The effective exponents
obtained from the total incremental times are $p=1.9$--$2.0$ between
adjacent steps and $p\approx2.0$ over the whole range $1600\to11000$,
consistent with $O(N^{2})$; over the same steps the standard
diagonalization gives $p\approx3.1$, consistent with $O(N^{3})$ and
with Table~\ref{tab:on3base}. (iii) The LOBPCG iteration count stays at
$544$--$575$, essentially independent of both the dimension
($N=1600$--$11000$) and the border size ($K=100$--$1000$)---precisely
the premise on which the $O(N^{2})$ count rests. The bordered Cholesky
update itself is negligible ($0.005$--$0.8$~s). (iv) The wall-clock
ratio (standard over incremental) grows nearly linearly with $N$:
$0.5$ at $N=1600$, $1.1$ at $3100$, $2.0$ at $5100$, and $4.5$ at
$11000$, with the crossover near $N\approx2700$. At small $N$ the ratio
is smaller than the ratio of operation counts because the incremental
side consists of memory-bound matrix--vector operations, whose
effective speed is lower than that of the matrix--matrix operations
underlying the standard diagonalization; since the exponents differ,
however, the ratio keeps growing with $N$, and extrapolating the
measured trend gives a factor of about $40$ at $N=10^{5}$. The
$O(N^{3})$ rediagonalization cost can therefore be avoided in
full-scale calculations that select tens of thousands of basis
functions.

Two remarks close this subsection. First, the success of the bordered
factorization alone does not certify linear independence: unless the
diagonal of the Schur complement is tested at every adoption,
$\min_r (L_{22})_{rr}^2 > s_{\rm cut}$, a numerically singular overlap
matrix passes silently and leads to a variational collapse with
spuriously deep energies---a behavior we confirmed in a separate check.
This test is the incremental counterpart of the removal of
zero-eigenvalue states performed by the canonical orthogonalization in
the standard diagonalization. Second, the $O(N^2)$ property presupposes that the LOBPCG
iteration count does not grow with $N$; this in turn relies on keeping
the overlap matrix well conditioned, which is the combined effect of the
unit-diagonal rescaling of Sec.~\ref{subsec:basis}, the Schur-diagonal
cut in the bordered update, and the trimming of Sec.~\ref{subsec:trim}.

\section{Results}
\label{sec:results}
\begin{figure}
\includegraphics[width=\columnwidth]{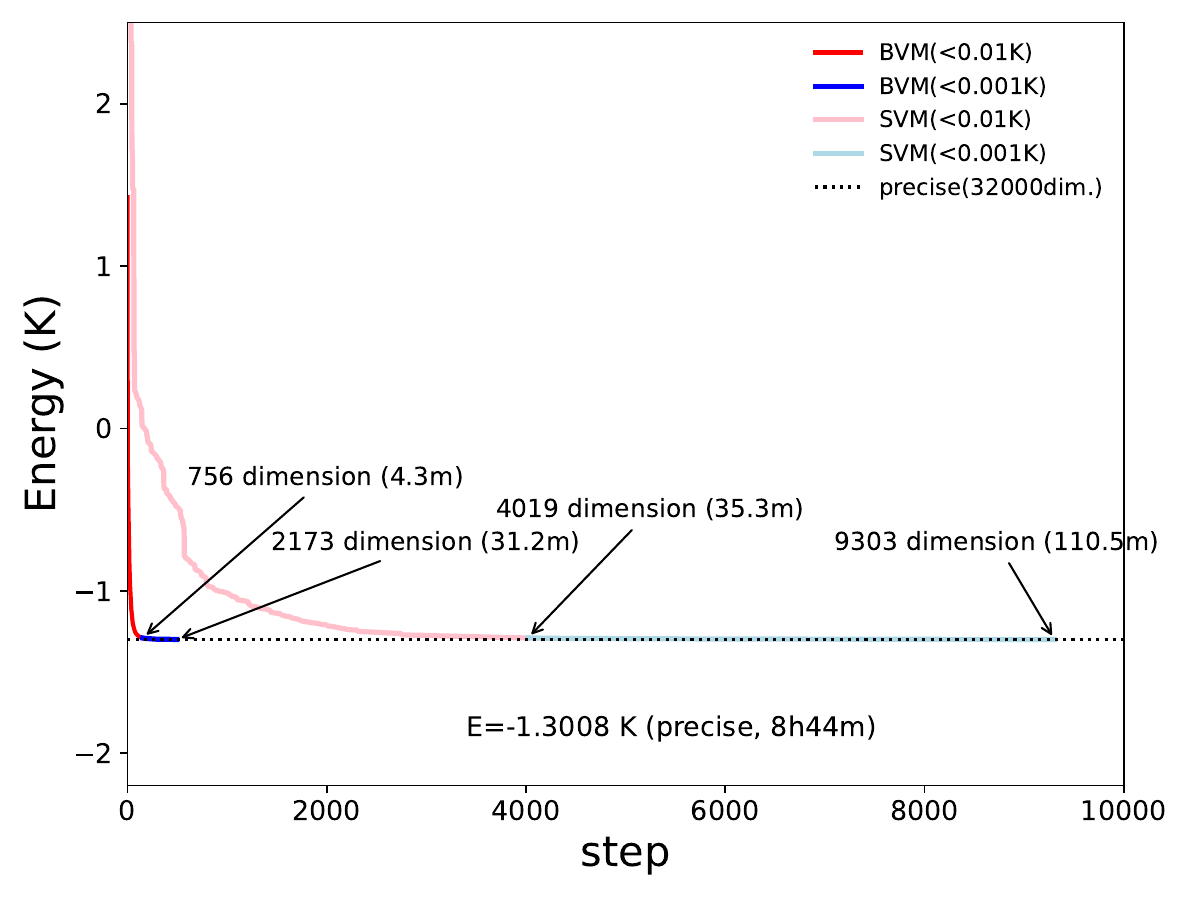}
\caption{Convergence of the variational energy $E$ for the $^4$He pentamer system
as a function of the optimization step. The BVM and SVM approaches are compared against
the precise reference energy $E_{\text{precise}} \approx -1.3008$\,K (black dotted line). 
Labels indicate the basis dimension and total execution 
time required to reach the accuracy criteria of $0.01$\,K and $0.001$\,K for each method.}
\label{fig:conv}
\end{figure}

Having validated the individual components in
Sec.~\ref{sec:validation}, we now apply the BVM to the full-scale
benchmark: the ground state of the $^4$He pentamer, computed on the
precomputed $32{,}000$-dimensional GEM basis, in comparison with the
conventional SVM.

Fig.~\ref{fig:conv} shows the convergence of the
variational energy as the basis set is progressively
enlarged. The horizontal axis represents the
optimization step, while the vertical axis shows the
resulting energy in Kelvin (K). Three calculations are
compared: the BVM, the SVM, and the reference energy
$E = -1.3008$\,K (black dotted line), obtained by full
diagonalization of the $32{,}000$-dimensional basis.

The set of $32{,}000$ candidate basis functions employed in
both the SVM and the
BVM is identical
to that used in the precise calculation represented by
the black dotted line in Fig.~\ref{fig:conv}. It is also
the same as that adopted in our preceding work, where
details are summarized in Table~II of Ref.~\cite{base}.
The Hamiltonian and overlap matrix elements associated
with these basis functions, which are constructed using
the GEM, are precomputed and
stored. Restricting the search to this finite candidate
set has three advantages. First, the reference solution
is unambiguous, as it is uniquely defined as the
converged value obtained by full diagonalization of the
$32{,}000$-dimensional basis. Second, the computational cost
per trial is reduced because the matrix elements are
reused rather than recomputed at each step. Third,
because this candidate set is constructed from the
geometric-progression Gaussian basis (grid GEM), the
resulting reference solution itself serves as a highly
accurate benchmark. On the other hand, unlike the conventional SVM, which samples new
functions from the continuous parameter space, the present approach
restricts the search to the grid; the slight loss of flexibility is the
trade-off accepted for the three advantages above.

In the SVM implementation used for the comparison in
Fig.~\ref{fig:conv}, we adopted a typical algorithm where
one basis function is added at each optimization step.
Specifically, a candidate basis function was selected
uniformly at random from the candidate set, which
initially consisted of the $32{,}000$ basis functions used in
the GEM precise calculation. The variational energy was
evaluated after augmenting the current basis with the
candidate. The candidate was accepted if the energy
reduction exceeded the current threshold, $\Delta E$
(initially $10^{-5}$\,K). Otherwise, another candidate was
randomly selected from the remaining set and tested in the
same manner. After 20 unsuccessful trials, $\Delta E$ was
reduced by one order of magnitude, thereby relaxing the
acceptance criterion.

In the BVM calculation shown in Fig.~\ref{fig:conv}, a
pool of $N_{\rm pool}=1000$ basis-function candidates was first generated
randomly from the set of $32{,}000$ candidate basis
functions, and five-element tuples were subsequently
selected from this pool. This pool--tuple hierarchy
reduces the number of candidate combinations
from ${}_{32000}C_{5}$ to ${}_{1000}C_{5}$. Even then, the
number of possible five-element combinations reaches
${}_{1000}C_{5} \approx 8.3 \times 10^{12}$, rendering an
exhaustive search impractical. Instead, 500 randomly
generated combinations were initially evaluated to
construct the teacher set for the GPR model. Based on
the resulting surrogate model, 500 additional
combinations predicted to yield low energies were
proposed and explicitly diagonalized. The newly evaluated
data were then incorporated into the teacher set, and
this procedure was repeated four times, resulting in a
total of $2{,}500$ explicitly evaluated combinations per
optimization step, where 500 were randomly selected and
$2{,}000$ were selected via the GPR surrogate model---the
same schedule as validated in Sec.~\ref{sec:val_main}. The 500 explicit diagonalizations of each round were distributed over
500 nodes through MPI, one tuple per node; since both the random
generation of trial tuples and their incremental evaluations are mutually
independent, this stage parallelizes almost ideally; with the residual
losses coming mainly from the progress management on the master
node---task assignment, collection of the results, and the determination
of the current minimum---and from the MPI communication, the measured
speedup is somewhat above 400, close to the ideal value of 500. The five-element tuple yielding
the lowest energy was adopted at that optimization step.

Furthermore, every 20 optimization steps, the trimming
procedure of Sec.~\ref{subsec:trim} was applied to remove
nearly linearly dependent basis functions and reduce the basis
dimension; this procedure accelerated the calculations
and saved memory.

All diagonalizations were carried out on the supercomputer
Fugaku, using 48 cores of a single node through OpenMP
parallelization. The full diagonalization of the
$32{,}000$-dimensional basis, which provides the precise
GEM reference energy, was performed using a standard
LAPACK routine. In contrast, the diagonalizations
required at each optimization step in the BVM and the SVM were performed using the $O(N^2)$
incremental diagonalization of Sec.~\ref{subsec:increment}. This
method reuses the stored Cholesky factorization and ground state of the
already accepted basis, thereby avoiding rediagonalization of the entire
enlarged basis from scratch at every step.
The conventional SVM is
essentially a single-node algorithm, and the basis-function
selection at each optimization step is not parallelized.
In contrast, the GPR component
of the BVM can be
readily parallelized. In particular, the explicit
evaluations of candidate tuples required for generating
teacher data can be distributed over a large number of
nodes. The BVM calculations shown in Fig.~\ref{fig:conv}
were performed using 501 nodes of the supercomputer Fugaku,
with one node assigned to overall control and the remaining
500 nodes dedicated to teacher-data generation and
related tasks.

We now turn to what Fig.~\ref{fig:conv} shows. A
calculation requiring the full $32{,}000$-dimensional basis
in the standard grid GEM converges, in the BVM, already
with a basis of only 756 dimensions at the three-digit
($0.01$\,K) accuracy, and still with only $2{,}173$ dimensions
at the four-digit ($0.001$\,K) accuracy. Since the matrix
storage grows as the square of the basis dimension, this
corresponds to a reduction of the memory footprint by
99.94\% and 99.54\%, respectively, relative to the full
$32{,}000$-dimensional calculation. In the present benchmark
this saving is not realized as such, because the full
$32{,}000$-dimensional matrices are precomputed and kept in
storage; in an actual application, however,
the required matrix elements are computed
within the low-dimensional setting, as they are needed,
and the memory actually used is that of the
accepted basis, which holds only a small number of functions.

In practical many-body calculations, such as those for
nuclei and hadronic systems, an accuracy of three to four
significant digits is usually sufficient. At the three-
digit level of accuracy (within $0.01$\,K), the BVM reaches the reference
solution in 4.3 min, even with the 501-node configuration.
In contrast, the conventional single-node SVM requires 35.3 min, while the
complete grid-GEM
diagonalization of the full $32{,}000$-dimensional basis
requires 8 h 44 min. At the four-digit level (within $0.001$\,K), the BVM reaches the reference solution in
31.2 min, whereas the SVM requires $9{,}303$ basis functions
and 110.5 min to attain the same accuracy.
Therefore, the present method is not
only faster than the conventional approach but also
provides results based on a more thoroughly optimized basis
of fewer functions.

Admittedly, one node against 501 is not a like-for-like comparison. The
comparison is nevertheless the practically relevant one: the SVM timing
represents the algorithm as it has actually been employed in the past
calculations of the authors and their collaborators, whereas the BVM
was designed from the outset for massively parallel execution. We
note that a purely random search of the type used in the validation of
Sec.~\ref{sec:validation} is itself a natural parallel extension of the
SVM and therefore a direct point of comparison; the comparison with the
GPR-guided search is presented in Table~\ref{tab:poolscan} below.

On a massively parallel supercomputer such as Fugaku,
which comprises more than $150{,}000$ nodes in total, the
allocation of 501 nodes is relatively modest, corresponding to only
about 1/300 (0.3\%) of the entire system resources. Moreover, computing
the matrix elements of the $32{,}000$ candidate basis functions used in this
verification also takes only about ten minutes on the same 500 nodes,
which for the five-body system is roughly comparable to the time of the
search itself. Ample room is thus left for extending the calculations to
six- and seven-body systems.

\begin{table}[!t]
\centering
\caption{Dependence on the number of GPR rounds $R$ per optimization
step in the full-scale setting with $N_{\rm pool}=1000$, with the
trimming switched off. The second column gives the number of tuples
evaluated per step, $500+500R$. Each entry is a single run with a
common random seed; $N$ is the reached basis dimension and $t$ the
wall-clock time in minutes.}
\label{tab:rscan}
\begin{tabular}{cccccc}
\hline
 & & \multicolumn{2}{c}{$0.01$\,K} & \multicolumn{2}{c}{$0.001$\,K} \\
$R$ & tuples/step & $N$ & $t$ [min] & $N$ & $t$ [min] \\
\hline
$1$ & $1000$ & $830$ & $1.3$ & $2555$ & $7.3$ \\
$3$ & $2000$ & $745$ & $1.9$ & $2420$ & $11.6$ \\
$5$ & $3000$ & $750$ & $3.0$ & $2390$ & $17.2$ \\
$7$ & $4000$ & $730$ & $3.8$ & $2355$ & $22.8$ \\
$9$ & $5000$ & $790$ & $5.9$ & $2395$ & $30.6$ \\
\hline
\end{tabular}
\end{table}

\begin{table*}[!t]
\centering
\caption{Dependence on the pool size $N_{\rm pool}$ and role of the
trimming step (Sec.~\ref{subsec:trim}) in the full-scale setting, for
the GPR-guided (GPR) and the purely random (RND) search: the basis
dimension $N$ and the wall-clock time $t$ (in minutes) at which the
$0.001$K criterion is reached (see text for the common setting).}
\label{tab:poolscan}
\begin{tabular}{ccccccccc}
\hline
 & \multicolumn{4}{c}{GPR} & \multicolumn{4}{c}{RND} \\
 & \multicolumn{2}{c}{trimming on} & \multicolumn{2}{c}{trimming off}
 & \multicolumn{2}{c}{trimming on} & \multicolumn{2}{c}{trimming off} \\
$N_{\rm pool}$ & $N$ & $t$ [min] & $N$ & $t$ [min]
               & $N$ & $t$ [min] & $N$ & $t$ [min] \\
\hline
$100$  & $2942$ & $54.9$ & $3230$ & $22.2$ & $3263$ & $58.9$ & $3940$ & $28.1$ \\
$500$  & $2231$ & $32.3$ & $2485$ & $14.7$ & $2857$ & $44.7$ & $3415$ & $21.7$ \\
$1000$ & $2173$ & $31.2$ & $2390$ & $14.4$ & $2836$ & $46.2$ & $3385$ & $21.3$ \\
$3000$ & $2177$ & $33.5$ & $2505$ & $18.5$ & $2802$ & $44.6$ & $3420$ & $23.3$ \\
$5000$ & $2204$ & $39.6$ & $2515$ & $22.7$ & $2806$ & $46.1$ & $3495$ & $27.1$ \\
\hline
\end{tabular}
\end{table*}

Since the GPR proposals carry a significant computational cost, the
number of propose--evaluate--update rounds per optimization step,
denoted $R$, is a critical parameter of the method.
Table~\ref{tab:rscan} examines this
dependence in the full-scale setting, with the trimming switched off so
that the effect of the GPR search is seen in its pure form. Each row of
the table corresponds to the number of rounds $R$. In each round the
GPR model proposes $500$ tuple candidates that are evaluated
explicitly, so that, together with the initial teacher set of $500$
random tuples, $500+500R$ tuples are evaluated per step (second
column). For each of the two accuracy criteria the table lists the
reached basis dimension $N$ and the wall-clock time $t$ (in minutes).
The reached dimension depends only weakly on $R$: at the $0.001$K
level it decreases gradually from $R=1$ to $R=7$, by about $8\%$;
considering that the uncertainty amounts to a few tens of dimensions
and that the value rises again at $R=9$, the dimension can be regarded
as essentially saturated around $R=3$--$5$. Since a smaller $R$ also
reduces the overall computation time, we use $R=4$ in the calculations
of this section, including Fig.~\ref{fig:conv}.

Table~\ref{tab:poolscan} examines the dependence on the pool size
$N_{\rm pool}$ and the role of the trimming step of
Sec.~\ref{subsec:trim} in the full-scale setting; in addition, it
provides the direct comparison of the GPR-guided search with the purely
random search (denoted GPR and RND, as in
Sec.~\ref{sec:validation}), the latter of which, as noted above, may
itself be regarded as a parallel extension of the conventional SVM.
The reason is as follows. At each optimization step the conventional
SVM tests randomly generated candidates one at a time against an
acceptance threshold, so that an adoption decision is typically reached
within a few to a few tens of trials; the random search instead
generates $2{,}500$ random tuples at once in a single step. It differs
in the acceptance rule---the tuple yielding the lowest energy is
adopted---but in most cases the adopted tuple also satisfies the usual
SVM acceptance criterion. It differs also in that five basis functions
are searched simultaneously rather than one; since the SVM does not in
itself require the search to proceed one function at a time, this is
the natural choice in a comparison with the BVM, which adopts
five-element tuples.

All runs share the full-scale setting of Fig.~\ref{fig:conv}, with
$2{,}500$ tuples evaluated per optimization step (an initial
teacher set of $500$ random tuples followed by four
propose--evaluate--update rounds of $500$); the random counterpart
replaces the GPR rounds by $2{,}500$ randomly generated tuples per step,
evaluated and adopted by exactly the same rule, so that the two searches
evaluate an identical number of tuples. Runs were performed for pool sizes
$N_{\rm pool}=100$--$5000$, with the trimming switched on and off. Each
row of the table corresponds to one pool size, and for each of the four
conditions the table lists separately the reached basis dimension $N$
and the wall-clock time $t$ at which the $0.001$K criterion is reached.
Each entry is a single run with a common random seed, so the individual
numbers carry a single-trial uncertainty of a few percent.

We first consider the dependence on the pool size. For the GPR-guided
search, enlarging the pool from $N_{\rm pool}=100$ to $1000$ reduces
the reached dimension substantially, from $2942$ to $2173$, and the
time from $54.9$ to $31.2$ min; enlarging it further no longer reduces
the dimension, while the time increases. A pool of about $1000$
candidates is therefore the operating point adopted for the full-scale
calculations, consistent with the saturation trend observed in
Sec.~\ref{sec:val_pool}. The random
search, in turn, does not improve beyond $N_{\rm pool}\approx500$: with
$2{,}500$ random samples per step, a larger pool cannot be exploited,
and both the reached dimension and the time remain essentially
unchanged.

We next consider the role of the trimming step. Switching trimming off
enlarges the final basis for both searches, by about $8$--$15\%$ for the
GPR-guided search and $16$--$25\%$ for the random one. Trimming thus
helps both sides, and appears to help the random search somewhat
more---consistent with the expectation that a search which adopts better
bases in the first place leaves less linear dependence for trimming to
remove.

Finally, the comparison with the purely random search confirms, in the
full-scale setting, the result of Sec.~\ref{sec:validation}: the
GPR-guided search reaches the criterion with a smaller basis (by
$10$--$23\%$) and in less time for every pool size, whether
or not trimming is applied. The time advantage stems from the smaller
accepted basis and from the higher efficiency of the adoption: per step
the random search is faster, having no surrogate overhead, but this is
outweighed by the smaller basis dimension---and hence the lower cost of
each diagonalization---and by the smaller number of steps required to
reach the target.

Table~\ref{tab:bvmruns} summarizes the performance of the BVM across ten independent trials,
where each trial corresponds to a unique sequence of
basis-function selection initiated with a different
random seed; trial~1 is the run shown in
Fig.~\ref{fig:conv}. The rows trial~1--10
identify the individual trials. The
subsequent columns list the required basis dimension and
the total wall-clock time to achieve two distinct accuracy
criteria (0.01\,K and 0.001\,K). As shown in Table~\ref{tab:bvmruns},
these independent trials yield closely consistent results:
at the 0.001K level the reached dimension is
$2170\pm22$ (range 2127--2209), a spread of about $1\%$,
and the wall-clock times lie within 28.1--32.6 min. The small
variation across trials demonstrates that the present
BVM robustly explores the optimized basis space,
consistently reaching an optimal solution with few basis functions regardless of
the initial random seed.

\begin{table}[h]
\centering
\caption{Basis dimensions and total wall-clock times of the BVM for
reaching specified accuracy levels, over ten independent trials in the
setting of Fig.~\ref{fig:conv}.}
\label{tab:bvmruns}
\begin{tabular}{lcccc}
\hline
trial & Dimension & Dimension & Time to & Time to \\
    & at $0.01$\,K & at $0.001$\,K & $0.01$\,K [min] & $0.001$\,K [min] \\
\hline
1 & 756 & 2173 & 4.3 & 31.2 \\
2 & 786 & 2175 & 4.3 & 30.4 \\
3 & 759 & 2160 & 4.3 & 30.4 \\
4 & 777 & 2160 & 4.9 & 31.0 \\
5 & 728 & 2152 & 4.2 & 30.5 \\
6 & 705 & 2127 & 3.9 & 28.1 \\
7 & 771 & 2195 & 4.4 & 31.3 \\
8 & 815 & 2209 & 5.1 & 31.2 \\
9 & 820 & 2161 & 5.0 & 32.6 \\
10 & 801 & 2183 & 5.0 & 31.2 \\
\hline
Mean & 772 & 2170 & 4.5 & 30.8 \\
Max  & 820 & 2209 & 5.1 & 32.6 \\
Min  & 705 & 2127 & 3.9 & 28.1 \\
\hline
\end{tabular}
\end{table}

\section{Summary and conclusions}
\label{sec:summary}

We have proposed the BVM, in which the basis functions of the
variational method are selected by Bayesian optimization. The basis is grown by
$K$-tuples drawn from a small random pool of candidates, so that
several functions are adopted together---an enlargement in groups that
can also capture the energy gain that strongly coupled basis functions
realize only in combination (Sec.~\ref{sec:intro}); a
Gaussian-process regression surrogate model with a Tanimoto kernel,
conditioned on all tuples evaluated so far, proposes the next candidates
through a lower-confidence-bound criterion; and every explicit
evaluation is carried out by the $O(N^2)$ incremental diagonalization,
which reuses the stored Cholesky factor and the ground state of the
accepted basis instead of solving the full eigenvalue problem from
scratch. An $O(N^2)$ trimming step continually removes basis functions
that have become nearly linearly dependent. All parts of the algorithm
were designed for massive parallelism, and the candidate
evaluations---the dominant cost---distribute over the nodes almost
ideally.

The ingredients were validated separately in Sec.~\ref{sec:validation}:
for an equal number of explicit evaluations the GPR-guided selection
reaches distinctly lower energies than a purely random search, robustly over
the tuple size $K$, the pool size $N_{\rm pool}$, and the
hyperparameters, whose influence is suppressed by the large evaluated
group of each round; the accuracy and the $O(N^2)$ scaling of the
incremental update were verified numerically on the benchmark matrices:
the energies agree with the standard diagonalization to within
$2\times10^{-10}$\,K, the iteration count is essentially independent of
the dimension, and the measured advantage over the $O(N^3)$
diagonalization grows nearly linearly with $N$---about a factor of $40$
at $N=10^{5}$ by extrapolation.

In the benchmark on the $^4$He pentamer with the $32{,}000$ precomputed
GEM basis functions (Sec.~\ref{sec:results}), the BVM reproduces the
reference energy of the full $32{,}000$-dimensional diagonalization to
within $0.01$\,K with only $705$ basis functions and to within
$0.001$\,K with $2{,}127$, corresponding to reductions of the matrix
memory by $99.95\%$ and $99.56\%$. On $501$ nodes of Fugaku this takes
$3.9$ and $28.1$~min, respectively, compared with $35.3$ and
$110.5$~min for the single-node SVM and $8$~h~$44$~min for the full
diagonalization. The outcome is insensitive to the random seed---the
reached dimension varies by about $1\%$ across independent trials---a
pool of about $1000$ candidates is a reasonable operating point, and
the GPR-guided search outperforms a purely random parallel
search---itself a natural parallel extension of the SVM---with an equal
number of evaluations for every pool size, reaching the criterion with
a $10$--$23\%$ smaller basis in less time, with and without trimming;
the trimming, in turn, keeps the final basis smaller for both searches
(switching it off enlarges the basis by $8$--$15\%$ for the GPR-guided
and $16$--$25\%$ for the random search).

The resources used remain modest for a five-body calculation---a
problem that has been regarded as a large-scale numerical
computation---amounting to only about $0.3\%$ of the entire Fugaku
system; moreover, in actual applications the matrix elements are
computed within the small accepted basis as they are needed, so that
the memory reduction is realized in full. Within the GEM, this makes
the six- and seven-body systems, which have been difficult with the
conventional procedure, the immediate next targets of precision
variational calculation. Moreover, the BVM
methodology applies Bayesian machine learning to the variational
parameters that describe the Hamiltonian and overlap matrices; it is
therefore not specific to the GEM, and its application to other
variational methods is also expected. Likewise, the
group adoption of the BVM should be advantageous wherever several basis
functions gain energy only in combination---as in the tensor
correlation of nuclei, where in the deuteron ground state the $L=0$ and
$L=2$ components couple strongly through the off-diagonal matrix
elements; the simultaneous adoption treats such off-diagonal components
directly, in contrast to a one-by-one acceptance.

In future work we will apply the method to six- and seven-body
systems. Applications to modern chiral effective-field-theory
interactions, which include the tensor force, will also be pursued.

\begin{acknowledgments}
This work was supported by the Japan Society for the Promotion of
Science (JSPS) KAKENHI, Grant Number JP25H01270, and by the ERATO TOMOE
project, Grant Number JPMJER2304. We used the supercomputer Fugaku at
the RIKEN Center for Computational Science (HPCI project ID: hp250328).
A part of the development was carried out on the supercomputer Miyabi
(the University of Tokyo and the University of Tsukuba), provided by
the Multidisciplinary Cooperative Research Program in the Center for
Computational Sciences, University of Tsukuba.
\end{acknowledgments}

\end{document}